\documentclass[11pt]{article}
\usepackage{amsmath,amssymb,amsthm,morefloats}
\usepackage{epsfig,psfrag,url}
\usepackage{graphicx}
\usepackage{verbatim}
\usepackage{subfigure}
\numberwithin{equation}{section}

\newcommand{\goto}{\rightarrow}
\newcommand{\bigo}{{\mathcal O}}

\newcommand{\half}{\frac{1}{2}}

\def\XXint#1#2#3{{\setbox0=\hbox{$#1{#2#3}{\int}$}
     \vcenter{\hbox{$#2#3$}}\kern-.5\wd0}}

\DeclareMathOperator{\diag}{diag}

\DeclareMathOperator{\imag}{Im}

\newenvironment{choices}{\left\{ \begin{array}{ll}}{\end{array}\right.}
\newcommand\when{&\text{if~}}
\newcommand\otherwise{&\text{otherwise}}

\newenvironment{mat}{\left[\begin{array}{ccccccccccccccc}}{\end{array}\right]}
\newcommand\bcm{\begin{mat}}
\newcommand\ecm{\end{mat}}

\newcommand{\bea}{\begin{eqnarray}}
\newcommand{\eea}{\end{eqnarray}}
\newcommand{\bean}{\begin{eqnarray*}}
\newcommand{\eean}{\end{eqnarray*}}
\newcommand{\ba}{\begin{array}}
\newcommand{\ea}{\end{array}}
\newcommand{\beqs}{\begin{equation*}\begin{split}}

\newcommand{\mc}{\mathcal}

\newtheorem{example}{Example}[section]

\newtheorem{remark}{Remark}[section]

\newtheorem{theorem}{Theorem}[section]

\newtheorem{definition}{Definition}[section]

\newtheorem{assumption}{Assumption}[section]

\long\def\symbolfootnote[#1]#2{\begingroup%
\def\thefootnote{\fnsymbol{footnote}}\footnote[#1]{#2}\endgroup}

\begin{document}
\title{A numerical dressing method for the nonlinear superposition of solutions of the KdV equation}
\author{Thomas Trogdon$^1$ \\
Courant Institute of Mathematical Sciences\\
New York University\\
251 Mercer St.\\
New York, NY 10012, USA\\
\phantom{.}\\
Bernard Deconinck\\
Department of Applied Mathematics\\
 University of Washington\\
Campus Box 352420\\
 Seattle, WA, 98195, USA \\}
\maketitle

\footnotetext[1]{Corresponding author, email: trogdon@cims.nyu.edu}

\begin{abstract}
In this paper we present the unification of two existing numerical methods for the construction of solutions of the Korteweg-de Vries (KdV) equation.  The first method is used to solve the Cauchy initial-value problem on the line for rapidly decaying initial data.  The second method is used to compute finite-genus solutions of the KdV equation.  The combination of these numerical methods allows for the computation of exact solutions that are asymptotically (quasi-)periodic finite-gap solutions and are a nonlinear superposition of dispersive, soliton and (quasi-)periodic solutions in the finite $(x,t)$-plane.  Such solutions are referred to as superposition solutions. We compute these solutions accurately for all values of $x$ and $t$.
\end{abstract}

\section{Introduction}

We consider the computation of solutions of the Korteweg-de Vries
\begin{align} \label{KdV}
q_t + 6qq_x + q_{xxx} = 0, ~~ (x,t) \in \mathbb R \times (0,T), ~~ T > 0,
\end{align}

\noindent with a particular class of step-like finite-gap initial data. 
For our purposes, $q_0(x)$ is said to be a step-like finite-gap function if
\begin{align*}
\left| \int_{0}^{\pm \infty} \left| \frac{d^n}{dx^n} (q_0(x) - q_\pm(x)) \right| (1 + |x|^m) dx \right| < \infty,
\end{align*}
for all non-negative integers $n$ and $m$ and some finite-gap potentials $q_\pm(x)$.  Finite-gap potentials $q_\pm(x)$ are those such that the operator $\partial_{xx} + q_\pm(x)$ admits a Bloch spectrum that consists of a finite number of intervals and the solution of \eqref{KdV} with $q_\pm$ as an initial condition is a finite-gap (or finite-genus) solution \cite{TheoryOfSolitons}.  In other words, $q_0(x)$ and its derivatives approach finite-gap potentials faster than any power, both as $x\rightarrow \infty$ and $x\rightarrow -\infty$. Recently, the existence and uniqueness of solutions for the KdV equation with this type of initial data was discussed for the case where the finite spectral bands associated with $q_\pm(x)$ either agree or are completely disjoint \cite{teschl-finite-gap}. It is shown there that the solution of the KdV equation satisfies

\begin{align}\label{sl-fg-time}
\left| \int_{0}^{\pm \infty} \left| \frac{d^n}{dx^n} (q(x,t) - q_\pm(x,t)) \right| (1 + |x|^m) dx\right| < \infty,
\end{align}
for all time.

\begin{remark}
The analysis in \cite{teschl-finite-gap} incorporates more general solutions then the numerical method discussed here.  We treat the case when the spectral bands of $q_\pm(x)$ coincide. For this reason we trade the term \emph{step-like finite-gap solution} in \cite{teschl-finite-gap} for \emph{superposition solution}.
\end{remark}

The results of \cite{teschl-finite-gap} present a significant step forward  in the analysis of the KdV equation.  Traditionally, the analysis proceeds in the Schwartz space ($q_\pm(x) = 0$) (for the whole line problem) or towards the construction of finite-genus solutions ($q_+(x) = q_-(x)$ and $q_0(x) = q_+(x)$) (the periodic or quasi-periodic problem).  Thus, the results in \cite{teschl-finite-gap} are a generalization of both the inverse scattering transform for rapidly decaying initial data \cite{ablowitz-segur-book,AblowitzClarksonSolitons} and of the analysis on Riemann surfaces for the construction of finite-genus solutions \cite{Dubrovin,TheoryOfSolitons}. In a similar way, the numerical approach we present for the construction of superposition solutions is a unification of existing numerical methods for the computation of rapidly decaying initial data and of finite-genus solutions.  The authors are not aware of any other existing method to compute superposition solutions.

The first method of two methods involved in the unification is used to compute solutions of the \emph{Cauchy initial-value problem on the line for rapidly decaying initial data} (IVP) \cite{TrogdonSOKdV}.  The second method is used to compute finite-genus solutions of the KdV equation. The approach we follow is based on a Riemann-Hilbert approach, as presented in \cite{TrogdonFiniteGenus}. A thorough discussion of the finite-genus solutions of the KdV equation is presented there as well. Our approach for computing the finite-genus solutions in \cite{TrogdonFiniteGenus} relies on a Riemann-Hilbert formulation, and is substantially different from the now standard approach of computing on Riemann surfaces, due to Bobenko and collaborators (using Schottky uniformization) \cite{belokolos1}, and Deconinck, Klein, van Hoeij, and others (using an algebraic curve representation of the Riemann surface), see \cite{deconinckpatterson} and \cite{frauendienerklein}, for instance. All the numerical approaches, both ours and the classical ones, rely on the theoretical work reviewed in \cite{TrogdonFiniteGenus} due to Its and Matveev \cite{itsmatveev1, itsmatveev2}, Novikov \cite{novikov} and Dubrovin \cite{dubrovin75}, McKean and van Moerbeke \cite{mckeanvanmoerbeke}, and others. An overview of the techniques used is presented in \cite{DubrovinTheta}, and a historical perspective can be found in \cite{matveev}.
  
We combine the approaches of \cite{TrogdonFiniteGenus} and \cite{TrogdonSOKdV}, and we show the evolution of solutions that are a nonlinear combination of finite-genus solutions and solutions of the IVP.  Despite the dispersive nature and quasi-periodicity of the solutions we are able to approximate them uniformly for all $x \in \mathbb R$ and $t\geq 0$. To combine the two approaches we use the dressing method (Section~\ref{Section:Dressing}, see also \cite[p.~221]{FokasUnified} and \cite{Dressing,ZakharovDressing}) as applied to the KdV equation.  This method allows us immense flexibility in the construction of solutions, in addition to providing a clear definition of the concept of \emph{nonlinear superposition}.  Following the classical works \cite{ablowitz-segur-book,TheoryOfSolitons} we begin with the spectral analysis of the time-independent Schr\"odinger equation:
\begin{align}\label{schrodinger}
-\Psi_{xx} - q(x,t) \Psi = \lambda \Psi,~~ k^2 = \lambda.
\end{align}
If $q$ solves \eqref{KdV} the spectrum of the operator $-\partial_{xx} - q(x,t)$ is independent of $t$.

Previous results have performed computation in the spectral $k$-plane when solving the IVP and in the $\lambda$-plane when constructing finite-genus solutions.  We show in Section~\ref{Section:Mapping} that the finite-genus solutions may be computed in the $k$-plane.  Therefore the dressing method may be applied directly in the $k$-plane. We present our numerical results in Section~\ref{Section:Results}.

\subsection{The solution of the initial-value problem with decay at infinity}

The dispersive nature of solutions of the IVP is highlighted in \cite{TrogdonSOKdV}.  A highly oscillatory dispersive tail moves with large velocity in the negative-$x$ direction.  This fact makes the approximation of solutions of the IVP difficult with traditional numerical methods.  The method in \cite{TrogdonSOKdV} derives it efficacy from the inverse scattering transform \cite{ablowitz-segur-book} and the Deift and Zhou method of nonlinear steepest descent \cite{DeiftZhouAMS}.  The solution of the IVP can be expressed in terms of the solution of a \emph{matrix Riemann--Hilbert problem} (RHP). Given an oriented contour $\Gamma$, an RHP poses the task of finding a sectionally analytic function $\Phi(k): \mathbb C \setminus \Gamma \goto \mathbb C^{m\times 2}$, depending on the parameters $x$ and $t$, such that
\begin{align*}
\lim_{\overset{z \goto k}{\text{ left of } \Gamma}} \Phi(x,t,z) = \left(\lim_{\overset{z \goto k}{z \text{ right of } \Gamma}} \Phi(x,t,z) \right) J(x,t,k), ~~ J(x,t,k) : \Gamma \goto \mathbb C^{2\times 2}.
\end{align*}
If $m=1$ we use  $\lim_{|k| \goto \infty} \Phi(k) = [1,1]$ and if $m =2$, $\lim_{|k| \goto \infty} \Phi(k) = I $.  Of course, the sense in which limits exist needs to be made precise, but this is beyond the scope of this paper, see \cite{zhou-RHP}.  We use the notation
\begin{align*}
\Phi^+(x,t,k) = \lim_{\overset{z \goto k}{z \text{ left of } \Gamma}} \Phi(x,t,z), ~~ \Phi^-(x,t,k) = \lim_{\overset{z \goto k}{z \text{ right of } \Gamma}} \Phi(x,t,z).
\end{align*}

  The RHP that appears in the solution of the IVP is oscillatory in the sense that $J(x,t,k)$ contains oscillatory factors.  Specifically, the RHP is of the form
\begin{align}\label{FokasLineRHP}
\Phi^+(x,t,k) &= \Phi^-(x,t,k) J(x,t,k),~~s \in \mathbb R, ~~\Phi(x,t,\infty) = [1,1],\\
J(x,t,k) &= \begin{mat}1 - \overline{\rho(\bar k)}\rho(k)  &  - \overline{\rho(\bar k)} e^{-2ikx - 8ik^3t}\\
\rho(k) e^{2ikx + 8ik^2t}& 1 \end{mat}.\notag
\end{align}
Once this is solved for $\Phi: \mathbb C \setminus \mathbb R \goto \mathbb C^{1\times 2}$ the solution $q(x,t)$ is found via
\begin{align}\label{Nonlinear-reconstruct}
q(x,t) = 2 i \lim_{|k| \goto \infty} k \partial_x\Phi_{1}(x,t,k),
\end{align}
where the subscript denotes the first component and $\rho(k)$ is the reflection coefficient that is computed accurately based on the initial condition \cite{TrogdonSOKdV}.  Note than when solitons are present in a solution of the KdV equation they manifest themselves as poles in the associated RHP.  Each soliton is uniquely specified by a pole $\kappa_j$ on the imaginary axis and a norming constant $c_j$.  In \cite{teschl,TrogdonSOKdV} it is shown how to remove these poles at the expense of introducing small contours on the imaginary axis.

\begin{figure}[t]
\centering
\includegraphics[width=.7\linewidth]{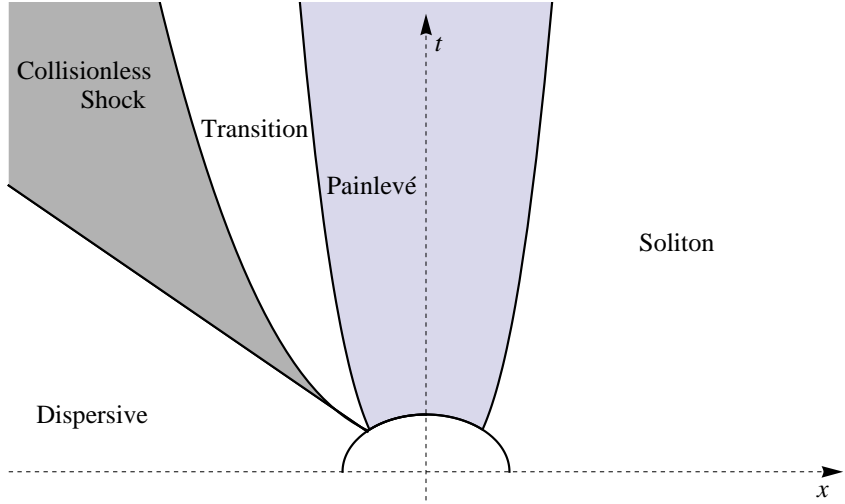}
\caption{\label{Figure:Regions} The different deformation regions of the KdV equation: the dispersive region, the collisionless shock region, the Painlev\'e region and the soliton region.}
\end{figure}

 The RHP can be deformed in much the same way as a contour integral so that oscillations turn to exponential decay.  The RHP is isolated near stationary phase points in the sense that the jump matrix is close to the identity matrix away from these stationary phase points.  The deformed RHP is solved approximately in terms of known functions. This is the essence of the method of nonlinear steepest descent.  An adaptation of it along with a numerical method for RHPs \cite{SORHFramework} is used to solve the RHP that arises in the solution of the IVP.  See Section~\ref{Section:Results} for plots of a numerical solution of the KdV equation obtained using this method.  The deformation required  to compute the solution varies as $x$ and $t$ vary.  We divide the $(x,t)$-plane into regions, guided by the classical asymptotic analysis \cite{AblowitzSegurSolution,deift-zhou-ven}.  Five regions exist; see Figure~\ref{Figure:Regions}.

It was noted in \cite{TrogdonSOKdV} that the computation of the solution of the KdV equation for moderate time can be completed without the use of the collisionless shock and transition regions.  More precisely, the dispersive region and the Painlev\'e region can be made to overlap up to some finite time $t^*$.  In this paper we show numerical results only for moderate time and we leave out the details of the deformations for the collisionless shock and transition regions.

Before we proceed with a discussion of the deformations we consider how poles in the RHP affect its definition.  It was shown in \cite{TrogdonSOKdV} (see also \cite{teschl}) that $\Phi$ can be redefined so that it solves
\begin{align*}
\Phi^+(x,t,k) &= \begin{choices} \Phi^-(x,t,k)J(x,t,k), \when  k \in \mathbb R,\\
\\
 \Phi^-(x,t,k)P_{j+}(x,t,k), \when  k \in A_j^+,\\
\\
\Phi^-(x,t,k) P_{j-}(x,t,k), \when k \in A_j^-,\end{choices}\\
\Phi(x,t,\infty) &= \begin{mat} 1 & 1\end{mat},\\
\end{align*}
where $A^-_j$($A^+_j$) are circular contours surrounding  $-\kappa_j$($+\kappa_j)$ with (counter-)clockwise orientation and 
\begin{align*}
P_{j+}(x,t,k) &= \begin{mat} 1 &0  \\ -c_j e^{\theta(\kappa_j)}/(k-\kappa_j) & 1 \end{mat},~~~ P_{j-}(x,t,k) = \begin{mat} 1 &  -c_j e^{\theta(\kappa_j)}/(k+\kappa_j) \\0 & 1 \end{mat},\\
\theta(k) &= 2ikx + 8ik^3t.
\end{align*}

\subsubsection{The dispersive region}

\begin{figure}[t]
\centering
\includegraphics[width=.8\linewidth]{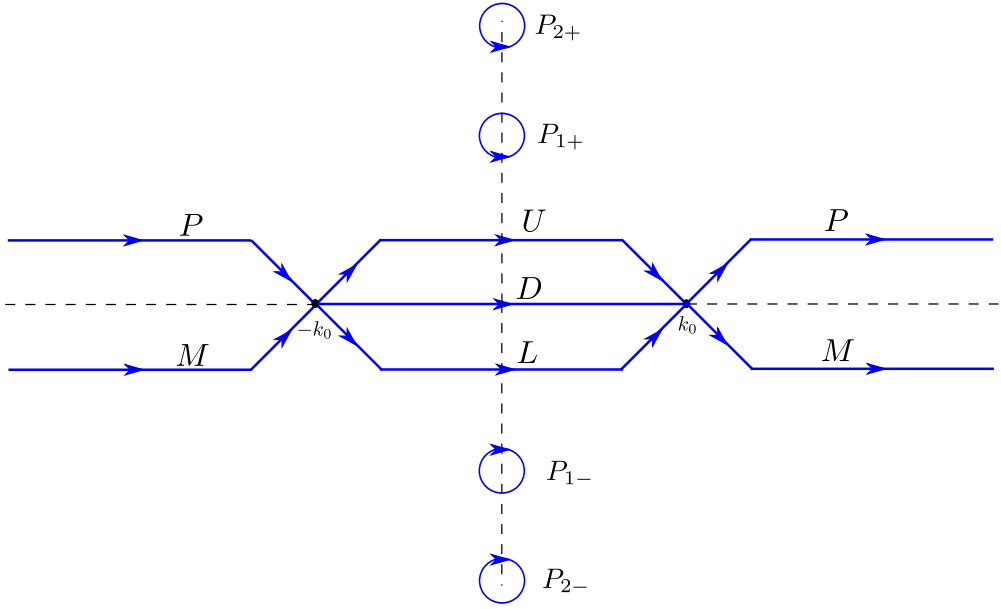}
\caption{\label{Figure:Phid1} The first deformation in the dispersive region.}
\end{figure}

The dispersive region is defined for $|-x/(12t)| < c_d$ for some constant $c_d$.  We introduce two algebraic factorizations of the jump matrix $J(x,t,k)$:
\begin{align*}
J(x,t,k) &= M(x,t,k)P(x,t,k),\\
M(x,t,k) &= \begin{mat} 1 & -\overline{\rho(\bar k)}e^{-\theta(k)} \\ 0 & 1\end{mat}, ~~P(x,t,k) = \begin{mat} 1 & 0 \\ \rho(k)e^{\theta(k)} & 1 \end{mat},\\
G(x,t,k) &= L(x,t,k)D(k)U(x,t,k), ~~ L(x,t,k)=\begin{mat} 1 & 0 \\ \rho(k)e^{\theta(k)}/(1-\rho(k) \overline{\rho(\bar k)}) & 1 \end{mat}, \\
D(k) &= \begin{mat} 1- \rho(k) \overline{\rho(\bar k)} & 0 \\ 0  & 1/(1-\rho(k) \overline{\rho(\bar k)}) \end{mat},~~ U(x,t,k) = \begin{mat} 1  & - \overline{\rho(\bar k)} e^{-\theta(k)}/(1 - \rho(k) \overline{\rho(\bar k)}) \\ 0 & 1 \end{mat}.
\end{align*}
Through the process known as lensing \cite[p.~192]{DeiftOrthogonalPolynomials} this RHP may be deformed to an RHP that passes along appropriate paths of steepest descent through the two stationary phase points $\pm k_0$ where $k_0 = \sqrt{-x/(12 t)}$.  This is illustrated in Figure~\ref{Figure:Phid1}.  

The off-diagonal entries of $P_{j\pm}$ may be  exponentially large depending on the values of $x$ and $t$.  Following the approach of \cite{teschl} we use a conjugation procedure to invert these exponentials when this is the case.  Define the index set
\begin{align*}
\mathcal K(x,t) = \left\{j: |c_j e^{\theta(\kappa_j)}| > 1 \right\},
\end{align*}
and the function
\begin{align*}
Q(k) = \begin{mat} \prod_{j\in \mathcal K(x,t)}  (k-\kappa_j)/(k+\kappa_j) & 0 \\ 0 & \prod_{j \in \mathcal K(x,t)} (k+\kappa_j)/(k-\kappa_j) \end{mat}.
\end{align*}
We define
\begin{align*}
\Phi_{1,d}(x,t,k) = \begin{choices} \Phi(x,t,k) \begin{mat} 1 & -(k-\kappa_j)/(C_je^{\theta(k_0)}) \\ C_j e^{\theta(k_0)}/(k-\kappa_j) & 0 \end{mat}Q(k), \when k \text{ is inside } A_j^+,\\
\\
\Phi(x,t,k) \begin{mat} 0 & -C_j e^{\theta(k_0)}/(k+\kappa_j) \\ (k+\kappa_j)/(C_j e^{\theta(k_0)}) & 1 \end{mat}Q(k), \when k \text{ is inside } A_j^-,\\
\\
\Phi(x,t,k)Q(k), \otherwise. \end{choices}
\end{align*}
It follows that this redefinition of $\Phi_{1,d}$ inside $A_j^\pm$ preserves analyticity away from the jump contour due to a removable singularity.  Define
\begin{align*}
N_{j+}(x,t,k) &= \begin{mat} 1 & -(k-\kappa_j)/(c_j e^{\theta(\kappa_j)}) \\ 0 & 1 \end{mat},~~~ N_{j-}(x,t,k) = \begin{mat} 1 &  0 \\-(k+\kappa_j)/(c_j e^{\theta(\kappa_j)}) & 1 \end{mat}.
\end{align*}
We compute the jumps that $\Phi_{1,d}$ satisfies:
\begin{align*}
\Phi_{1,d}^+(x,t,k) = \Phi_{1,d}^-(x,t,k) \begin{choices} Q^{-1}(k) N_{j\pm}(x,t,k) Q(k), \when k \in A_j^\pm \text{ and } j \in \mathcal K(x,t),\\
Q^{-1}(k) P_{j\pm}(x,t,k) Q(k), \when k \in A_j^\pm \text{ and } j \not\in \mathcal K(x,t),\\
Q^{-1}(k)J_1(x,t,k)Q(k), \otherwise, \end{choices}
\end{align*}
where $J_1$ represents the jump matrix for $\Phi$ in Figure~\ref{Figure:Phid1}.

\begin{figure}[tbp]
\centering
\subfigure[]{\includegraphics[width=.68\linewidth]{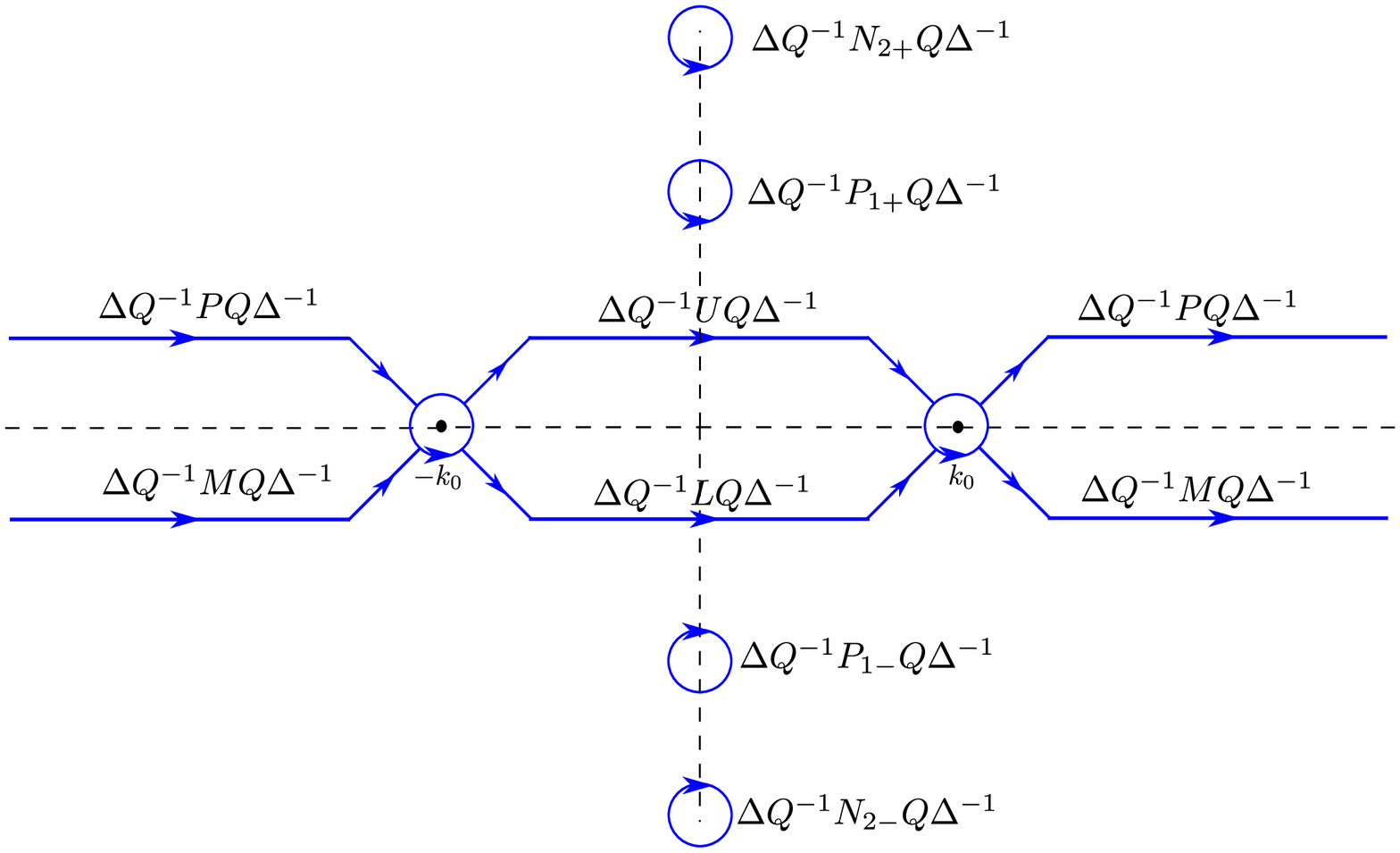}}
\subfigure[]{\includegraphics[width=.75\linewidth]{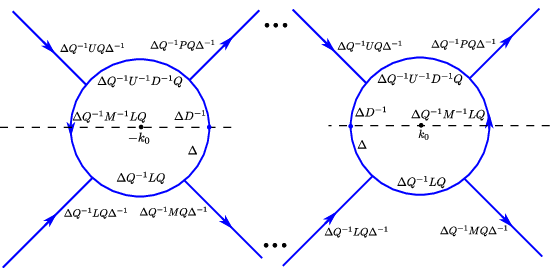}}
\caption{\label{Figure:Phi2d}  The jump contours and jump matrices of the RHP for $\Phi_{2,d}$. (a) The full contours in the case $\mathcal K(x,t) = \{2\}$.  (b) A zoomed view of the contours near the stationary phase points.
}
\end{figure}

Next we construct  parametrices, for numerical purposes.  The utility of these is made clear below.  Define 
\begin{align*}
\delta(k;k_0) = \exp\left( \frac{1}{2\pi i} \int_{-k_0}^{k_0} \frac{\log (1- \rho(s) \bar \rho(s))}{s-k} ds \right), ~~~ \Delta(k;k_0) = \diag(\delta(k;k_0),1/\delta(k;k_0)),
\end{align*}
so that $\Delta$ satisfies
\begin{align*}
\Delta^+(k;k_0) = \Delta^-(k;k_0) D(k), ~~ \Delta(\infty;k_0) = I.
\end{align*}
Note that $\Delta$ may be computed uniformly in the complex plane using the method in \cite{SOPainleveII}.  Next, define
\begin{align*}
H_{+k_0}(k) &= \begin{choices} \Delta^{-1}(k_0;k)D(k)U(x,t,k), \when \arg (k-k_0) \in (\pi/4,3\pi/4),\\
 \Delta^{-1}(k_0;k)D(k), \when \arg (k-k_0) \in (3\pi/4,\pi),\\
 \Delta^{-1}(k_0;k), \when \arg (k-k_0) \in (-\pi,-3\pi/4),\\
 \Delta^{-1}(k_0;k)L^{-1}(x,t,k), \when \arg (k-k_0) \in (-3\pi/4,-\pi/4),\\
 \Delta^{-1}(k_0;k)L^{-1}(x,t,k)M(x,t,k), \when \arg (k-k_0) \in (-\pi/4,0),
\end{choices}\\
H_{-k_0}(k) &= \begin{choices} \Delta^{-1}(k_0;k)D(k)U(x,t,k), \when \arg (k+k_0) \in (\pi/4,3\pi/4),\\
 \Delta^{-1}(k_0;k)D(k), \when \arg (k+k_0) \in (0,\pi/4),\\
 \Delta^{-1}(k_0;k), \when \arg (k+k_0) \in (-\pi/4,0),\\
 \Delta^{-1}(k_0;k)L^{-1}(x,t,k), \when \arg (k+k_0) \in (-3\pi/4,-\pi/4),\\
 \Delta^{-1}(k_0;k)L^{-1}(x,t,k)M(x,t,k), \when \arg (k+k_0) \in (3\pi/4,\pi)\cup (-\pi,-3\pi/4).
\end{choices}
\end{align*}
Let $r > 0$ and define
\begin{align*}
\Phi_{2,d}(x,t,k) = \Phi_{1,d}(x,t,k)\begin{choices}
Q^{-1}(k)H_{+k_0}^{-1}(k) \Delta^{-1}(k;k_0)Q(k), \when |k-k_0|<r,\\
Q^{-1}(k)H_{-k_0}^{-1}(k) \Delta^{-1}(k;k_0)Q(k), \when |k+k_0|<r,\\
Q^{-1}(k)\Delta^{-1}(k;k_0)Q(k), \otherwise .
\end{choices}
\end{align*}
The jump matrix for the RHP for $\Phi_{2,d}$ is shown in Figure~\ref{Figure:Phi2d}.  Note that $\Delta$ has (bounded) singularities at $\pm k_0$.  These deformations are chosen so that contours are located away from $\pm k_0$. 

\subsubsection{The Painlev\'e region}

\begin{figure}[tbp]
\centering
\includegraphics[width=.48\linewidth]{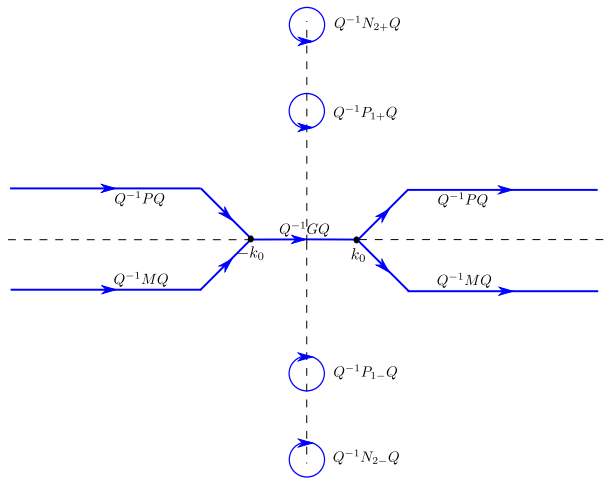}
\caption{\label{Figure:Phip1} The deformation in the Painlev\'e region when $x < 0$.  In this figure $\mathcal K(x,t) = \{2\}$.}
\end{figure}

The Painlev\'e region is defined for $|x| < c_p t^{1/3}$. This region overlaps with the soliton region up to $t^* = (12 c_d/c_p)^{-3/2}$.  Fortunately, the deformation of the RHP is simpler in the Painlev\'e region.  Under the assumption $|x| < c_p t^{1/3}$ it can be seen that the oscillations from $e^{\pm(2ikx+8ik^3t)}$ are controlled on $[-k_0,k_0]$.  We collapse the lens on $[-k_0,k_0]$ indicating that the $LDU$ factorization of the jump matrix is not needed in this region.  Furthermore, this implies that $\Delta$ is no longer needed for the deformation.  See Figure~\ref{Figure:Phip1} for the jump matrices and jump contours for the deformation in the Painlev\'e region when $x < 0$.  When $x > 0$ we use the deformation discussed in the next section.

\subsubsection{The soliton region}

\begin{figure}[tbp]
\centering
\subfigure[]{\includegraphics[width=.4\linewidth]{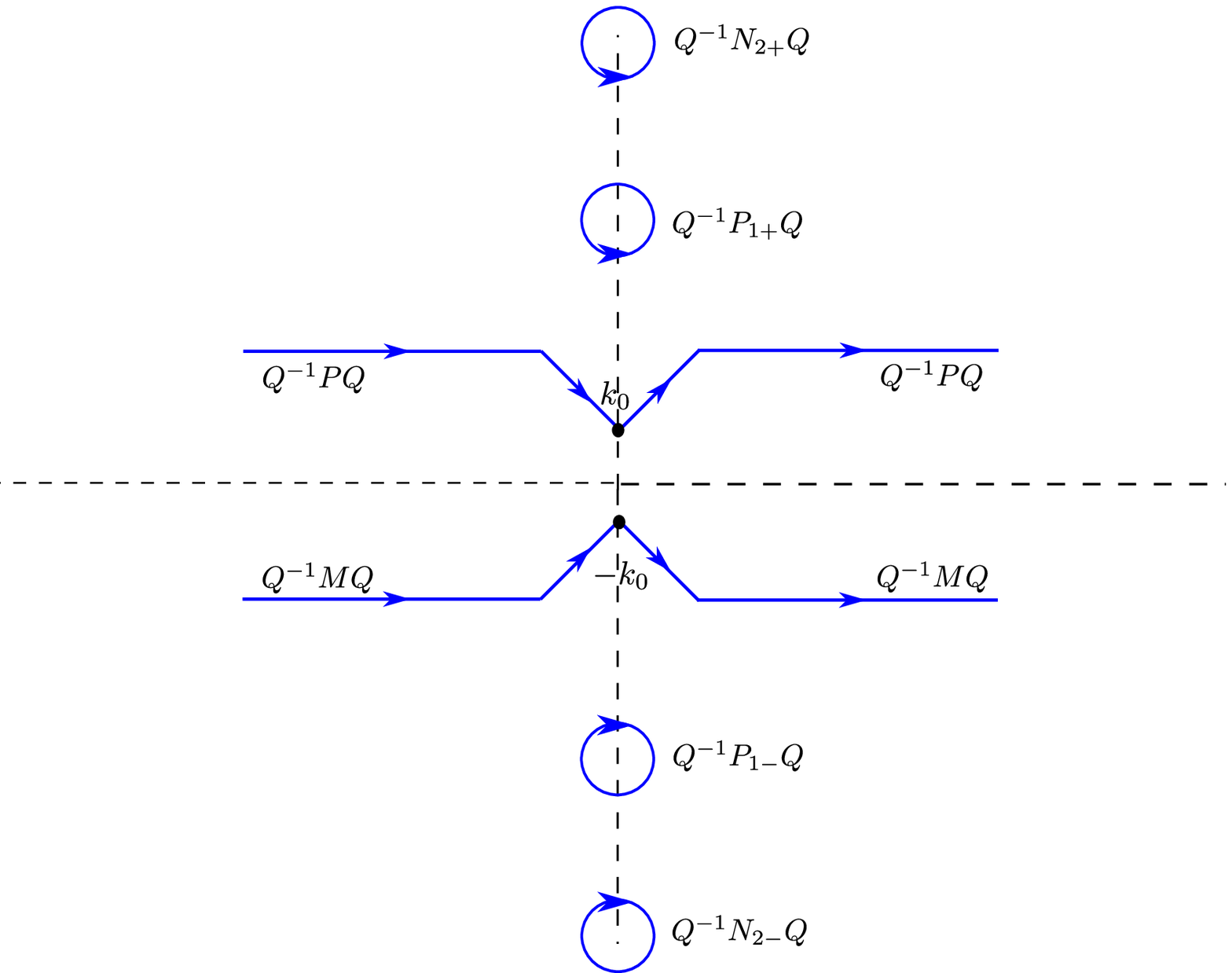}}
\subfigure[]{\includegraphics[width=.4\linewidth]{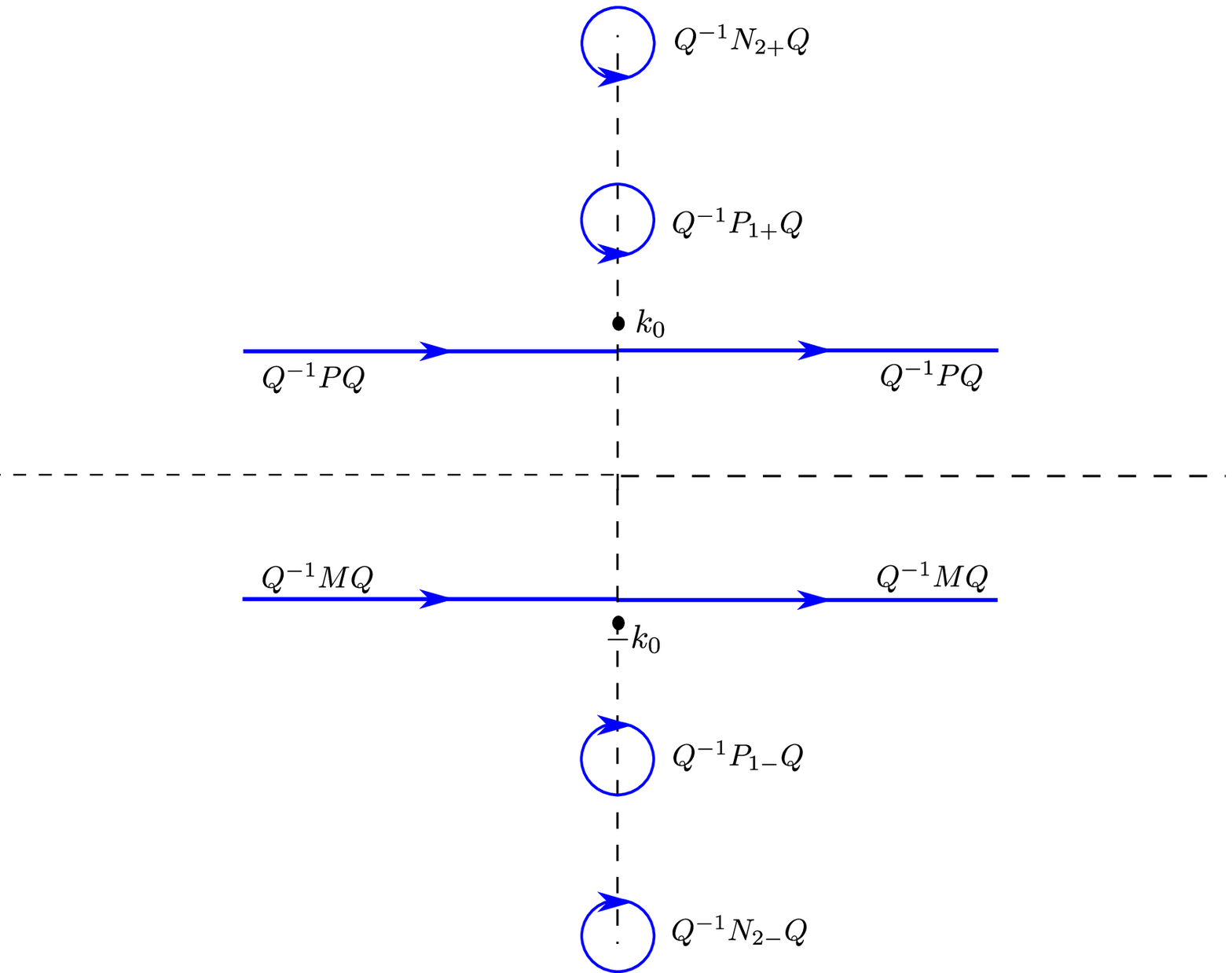}}
\caption{\label{Figure:Phis1} The deformation in the soliton region when $x < 0$.  In this figure $\mathcal K(x,t) = \{2\}$. (a) The deformation while the stationary phase point lies within the domain of analyticity for $\rho$. (b) The deformation when the stationary phase point is outside this domain of analyticity.}
\end{figure}

The deformation is further simplified in the soliton region ($x > c_pt^{1/3}$) and for $x> 0$ in the Painlev\'e region.  Note that for $x > 0$ the stationary phase points are purely imaginary and move away from the origin on the imaginary axis as $x$ increases.  It would be ideal to deform the contours through these points for all $x$ but this is not possible: for exponentially decaying initial data $\rho(k)$ is analytic only within a strip that contains the real line.  Thus, we deform though the stationary phase points until they leave a specified strip that contains the real line and is a subset of the domain of analyticity of $\rho$.  See Figure~\ref{Figure:Phis1} for the jump contours and jump matrices for the deformation in the soliton region.

\begin{remark}
We see that the strip $\{(x,t): x \in \mathbb R, 0 < t \leq t^*\}$ is entirely covered by these three regions.  Thus, by adjusting $c_d$ and $c_p$ we obtain a method that is accurate up to some finite time.  For arbitrarily large time, one must introduce the transition and collisionless shock regions, see Figure~\ref{Figure:Regions}.
\end{remark}

\subsection{Finite-genus solutions}

The finite-genus solutions of the KdV equation can be expressed in terms of the solution of an RHP as well.  Such an RHP was derived in \cite{TrogdonFiniteGenus}. Let $\Psi_\pm(x,t,\lambda)$ be solutions of \eqref{schrodinger} that satisfy $\Psi_\pm(x,t,\lambda) \sim e^{\pm i \lambda^{1/2} x \pm 4 i \lambda^{3/2} t}$ as $\lambda \goto \infty$. We restrict to the case where $q(x,t)$ solves \eqref{KdV} and $q(x,0)$ is a finite-gap potential.  In this case the spectrum of $-\partial_{xx} - q(x,t)$ is a subset of the real axis that consists of a finite number of finite-length intervals $\{(a_j^2,b_j^2)\}_{j=1}^g$ and one infinite interval $(a^2_{g+1},\infty)$.  We assume $a_1 = 0$ and $a_j < b_j < a_{j+1}$. It was shown in \cite{TrogdonFiniteGenus} that $\hat \Psi(x,t,\lambda) = [ \Psi_+(x,t,\lambda), \Psi_-(x,t,\lambda) ]$ satisfies
\begin{align*}
\hat \Psi^+(x,t,\lambda) = \hat \Psi^-(x,t,\lambda) \begin{mat} 0 & 1 \\ 1 & 0 \end{mat},  ~\lambda \in (a^2_{g+1},\infty) \cup \bigcup_{j=1}^g (a_j^2,b_j^2),\\
\hat \Psi(x,t,\lambda) = \begin{mat} e^{i\lambda^{1/2}x+4i\lambda^{3/2}} & e^{-i\lambda^{1/2}x-4i\lambda^{3/2}} \end{mat}( I + \bigo(\lambda^{-1/2})).
\end{align*}
Furthermore,
\begin{align*}
\hat \Phi(x,t,\lambda) = \hat \Psi(x,t,\lambda) \begin{mat} e^{-i\lambda^{1/2} x - 4i \lambda^{3/2} t} & 0 \\
0 & e^{i\lambda^{1/2} x + 4i \lambda^{3/2} t} \end{mat}
\end{align*}
satisfies
\begin{align}\label{almost-rhp}
\begin{split}
\hat \Phi^+(x,t,\lambda) &= \hat \Phi^-(x,t,\lambda) \begin{mat} 0 & 1 \\ 1 & 0 \end{mat},  ~\lambda \in (a_{g+1}^2,\infty) \cup \bigcup_{j=1}^g (a_j^2,b_j^2),\\
\hat \Phi^+(x,t,\lambda) &= \hat \Phi^-(x,t,\lambda) \begin{mat} e^{-2i\lambda^{1/2} x - 8i \lambda^{3/2} t} & 0 \\ 0 & e^{2i\lambda^{1/2} x +8i \lambda^{3/2} t} \end{mat}, ~ \lambda \in \bigcup_{j=1}^{g}(b^2_j,a^2_{j+1}),\\
\hat \Phi(x,t,\lambda) &= \begin{mat} 1 & 1 \end{mat}( I + \bigo(\lambda^{-1/2})).
\end{split}
\end{align}
It is shown in \cite{TrogdonFiniteGenus} that when viewed as an RHP, \eqref{almost-rhp} has non-unique solutions.  After a regularization procedure where choices are made, \eqref{almost-rhp} is converted into a problem with unique solutions. This regularized problem is solved numerically, and a numerical approximation of $q(x,t)$ is recovered from $\hat \Phi$ from the large $\lambda$ asymptotics.

The important aspect that we discuss below is that for $k^2 = \lambda$, we can express \eqref{almost-rhp} as RHP in the $k$-plane.  Thus computation in the $k$-plane can be used to produce finite-genus solutions.

\section{The Dressing Method}\label{Section:Dressing}

In this section, we discuss the construction of solutions of the KdV equation via the dressing method.  It follows from the inverse scattering transform (essentially, by construction) that $\Phi$ in \eqref{FokasLineRHP} satisfies the Jost equation
\begin{align}\label{x-jost}
-\Phi_{xx} + 2ik \Phi_x \sigma_3 - q(x,t) \Phi = 0, ~~ \sigma_3 = \begin{mat} 1 & 0 \\ 0 & -1 \end{mat}.
\end{align}
Furthermore, it is easy to check that $\hat \Phi$ (see \eqref{almost-rhp}) also satisfies this equation with $k$ replaced with $\lambda^{1/2}$. These functions satisfy a second equation determining their $t$-dependence \cite{ablowitz-segur-book,TheoryOfSolitons}:
\begin{align}\label{t-jost}
- \Phi_t + 4ik^3 \Phi \sigma_3 = (2 q(x,t)-4k^2)\left( \Phi_x - ik \Phi \sigma_3 \right) - q_x(x,t) \Phi.
\end{align}
Indeed \eqref{x-jost} and \eqref{t-jost} essentially make up the Lax pair for the KdV equation.  This is easily seen by writing $\Psi = \Phi e^{-(ikx + 4ik^3t) \sigma_3}$ and finding the differential equations solved by $\Psi$.  This produces the Lax pair in \cite[p.~70]{AblowitzClarksonSolitons}. This relationship is further explained by the dressing method.  Introduce the notation
\begin{align*}
e^{\alpha \hat \sigma_3} A = e^{\alpha \sigma_3} A e^{-\alpha \sigma_3}.
\end{align*}
We state the dressing method as a theorem.

\begin{theorem}\label{Theorem:Dressing}
Let $\Phi(x,t,k)$ solve the RHP
\begin{align*}
\Phi^+(x,t,s) = \Phi^-(x,t,s) e^{-\theta(x,t,s) \hat \sigma_3} V(s), ~~ s \in \Gamma,~~ \theta(x,t,s) = ikx+4ik^3t, ~~\Phi(x,t,\infty) = [1,1],
\end{align*}
where $\bar \Gamma = \Gamma$ (with orientation), $\det V(k) = 1$, $\overline{V(\bar k)} = V(-k)$ and $ V^{-1}(k) = \sigma_1 \overline{V(\bar k)} \sigma_1$ with
\begin{align*}
\sigma_1 = \begin{mat} 0 & 1 \\ 1 & 0 \end{mat}.
\end{align*}
Assume that the RHP has a unique solution that is sufficiently differentiable in $x$ and $t$ and that all existing derivatives are $\bigo(1/k)$ as $k \goto \infty$.  Define
\begin{align}\label{Q-reconstruct}
\begin{mat} Q(x,t) & Q(x,t) \end{mat} = 2i \lim_{k \goto \infty} k \partial_x \Phi(x,t,k) \sigma_3.
\end{align}
Then $\Phi(x,t,k)$ solves
\begin{align}\begin{split}\label{jost-pair}
-\Phi_{xx} + 2ik \Phi_x \sigma_3 &- Q(x,t) \Phi = 0,\\
- \Phi_t +  4ik^3 \Phi \sigma_3 &= (2 Q(x,t)-4k^2)\left( \Phi_x - ik \Phi \sigma_3 \right) - Q_x(x,t) \Phi,
\end{split}\end{align}
and $Q$ solves \eqref{KdV}.

\begin{proof}
We begin by establishing some symmetries of the solution. Let $\Phi$ be matrix-valued and tend to the identity matrix at infinity.  We show that this matrix RHP can be reduced to vector RHP.  The hypotheses of the theorem are sufficient to guarantee that such a matrix-valued solution is unique.  We show that the matrix problem can be reduced to that of a vector RHP.

 Define $\hat \Phi(k) = \overline{\Phi(-\bar k))}$.  Note that $\hat \Phi^+(k) = \overline{\Phi^+(-\bar k))}$ so that
\begin{align*}
\hat \Phi^+(k)  = \hat \Phi^-(k) \overline{V(-\bar k)} = \hat \Phi^-(k) V(k).
\end{align*}
Therefore by uniqueness, $\Phi(k) = \overline{\Phi(- \bar k)}$.  Expand $\Phi$ near $\infty$ using this symmetry:
\begin{align*}
\Phi(k) &= I + \Phi_1 k^{-1} + \Phi_2 k^{-2} + \bigo(k^{-3})\\
&= I - \bar \Phi_1k^{-1} + \bar \Phi_2 k^{-2} + \bigo(k^{-3}).
\end{align*}
Thus $\Phi_1$ is purely imaginary.  Next, define $\tilde \Phi(k) = \sigma_1 \overline{\Phi(\bar k)} \sigma_1$ and note that $\tilde \Phi^+(k) =  \sigma_1 \overline{\Phi^-(\bar k)} \sigma_1$.  We obtain
\begin{align*}
\tilde \Phi^+(k) = \sigma_1  \overline{\Phi^+(\bar k)} \overline{V^{-1}(\bar k)} \sigma_1 = \tilde \Phi^-(k) \sigma_1 \overline{V^{-1}(\bar k)} \sigma_1 = \tilde \Phi^-(k) V(k).
\end{align*} 
Thus $\Phi(k) = \sigma_1  \overline{\Phi(\bar k)}\sigma_1$.  Again, considering the series at infinity,
\begin{align*}
\Phi(k) &= I + \Phi_1 k^{-1} + \Phi_2 k^{-2} + \bigo(k^{-3})\\
&= I  + \sigma_1 \bar \Phi_1 \sigma_1  k^{-1} + \sigma_1 \bar \Phi_2 k^{-2} \sigma_1 + \bigo(k^{-3}).
\end{align*}
Therefore $\Phi_1 = \sigma_1 \bar \Phi_1 \sigma_1 = - \sigma_1 \Phi_1 \sigma_1$.  If
\begin{align*}
\Phi = \begin{mat} a & b \\ c & d \end{mat},
\end{align*}
then $a = -d$ and $c = -b$.  Let $\phi$ be the vector consisting of the sum of the rows of $\Phi$.  It follows that
\begin{align*}
\phi = [1,1] + \phi_1 k^{-1} + \bigo(k^{-2}),
\end{align*}   
where $\phi_1 \sigma_3 = Q(x,t)[1,1]$ for some scalar-valued function $Q$.  Thus the symmetries of the problem allow us to reduce it to a vector RHP, justifying \eqref{Q-reconstruct}. 

The fact that the RHP has a unique solution implies that the only solution that decays at infinity is the zero solution.  A straightforward but lengthy calculation shows that 
\begin{align*}
-\Phi_{xx} + 2ik \Phi_x \sigma_3 &- Q(x,t) \Phi,\\
 \Phi_t - 4ik^3 \Phi \sigma_3 &+ (2 Q(x,t)-4k^2)\left( \Phi_x - ik \Phi \sigma_3 \right) - Q_x(x,t) \Phi
\end{align*}
are solutions that decay at infinity.  Hence, we obtain \eqref{jost-pair}.  The compatibility condition of \eqref{jost-pair} implies $Q$ solves \eqref{KdV} as mentioned above.
\end{proof}
\end{theorem}

\subsection{A RHP on cuts}

With the ideas of the dressing method established, we consider the RHP
\begin{align}\label{delta-rhp}
\begin{split}
 \varphi^+(x,t,k) &=  \varphi^-(x,t,k) \begin{mat} 0 & -e^{-2ik x - 8ik^3 t} \\ e^{2ik x +8i k^{3} t} & 0 \end{mat}, ~~ k \in \bigcup_{j=1}^g \left( (-a_{j+1}, -b_j) \cup (b_j,a_{j+1}) \right),\\
 \varphi(x,t,k) &= \begin{mat} 1 & 1 \end{mat}( I + \bigo(1/k)),
\end{split}
\end{align}
where $0 < a_j < b_j < a_{j+1}$.  It follows that $q(x,t) = 2i\lim_{k\goto\infty} k \partial_x \varphi_1(x,t,k)$ must be a solution of the KdV equation. Below, we connect this solution to the finite-genus solutions and we superimpose this RHP on the RHP for the IVP to obtain dispersive finite-genus solutions in Section~\ref{Section:Results}.  In the remainder of this section we discuss the numerical solution of this RHP.

It is clear that \eqref{delta-rhp} is an oscillatory RHP.  Solutions of the RHP are more oscillatory as $|x|$ and $t$ increase.  We use the $g$-function mechanism \cite{Deift-gfun,zhou-notes} to remove these oscillations.  Consider the scalar RHP for $j=1,2,\ldots,g$:

\begin{itemize}
\item $\mc G^+(x,t,k) + \mc G^-(x,t,k) = 0$ for $ k \in (-a_{j+1},b_{j}) \cup (b_j, a_{j+1})$,
\item $\mc G^+(x,t,k) - \mc G^-(x,t,k) = -(2ikx+8ik^3t) + i\Omega_{j+}(x,t)$ for $k  \in (b_j,a_{j+1})$,
\item $\mc G^+(x,t,k) - \mc G^-(x,t,k) = -(2ikx+8ik^3t) + i\Omega_{j-}(x,t)$ for $k  \in (-a_{j+1},-b_j)$, and
\item $\mc G(x,t,k) = \bigo(k^{-1})$ as $k \goto \infty$.
\end{itemize}
Here  $\{\Omega_{j\pm}(x,t)\}_{j=1}^g$ are constants (with respect to $k$) to be determined.  It is straightforward to find a function $\mc G$ that satisfies the first three properties:
\begin{align*}
\mc G(x,t,k) = \frac{\sqrt{P(k)}}{2 \pi i} \sum_{j=1}^g \left (\int_{b_j}^{a_{j+1}} \frac{-(2isx+8is^3t) + i\Omega_{j+}(x,t)}{\sqrt{P(s)}^+} \frac{ds}{s-k} \right.\\
\left. +  \int_{-a_{j+1}}^{b_{j}} \frac{-(2isx+8is^3t) + i\Omega_{j-}(x,t)}{\sqrt{P(s)}^+} \frac{ds}{s-k} \right),
\end{align*}
where $P(k) = \prod_{j=1}^g \left [(k -b_j)(k-a_{j+1})(k+b_j)(k+a_{j+1}) \right]$. Here $\sqrt{P(k)}$ is taken to have branch cuts on the intervals $(b_j,a_{j+1})$ and $(-a_{j+1},-b_j)$ and the behavior $\sqrt{P(k)} \sim k^{2g}$ as $k \goto \infty$.  Furthermore, we define $\sqrt{P(k)}^+ = \lim_{\epsilon \goto 0^+} \sqrt{P(k+ i \epsilon)}$.  The set $\{\Omega_{j\pm}(x,t)\}_{j=1}^g$ is chosen so that $\mc G(x,t,k)~=~\bigo(k^{-1})$ as $k \goto \infty$.  Expanding $1/(s-k)$ in a Neumann series we find the $2g$ conditions:
\begin{align}\label{moment-conditions}
\begin{split}
 0 &= \sum_{j=1}^g \left (\int_{b_j}^{a_{j+1}} \frac{-(2isx+8is^3t) + i\Omega_{j+}(x,t)}{\sqrt{P(s)}^+} s^m ds +  \int_{-a_{j+1}}^{b_{j}} \frac{-(2isx+8is^3t) + i\Omega_{j-}(x,t)}{\sqrt{P(s)}^+} s^m \right),\\
m &= 0,1,\ldots 2g-1.
\end{split}
\end{align}
We obtain a linear system for $\{\Omega_{j\pm}(x,t)\}_{j=1}^g$.  The ideas from  \cite{TrogdonFiniteGenus} are adapted easily to show that this linear system is uniquely solvable.  Furthermore, it is demonstrated in \cite{TrogdonFiniteGenus} how to compute all integrals that appear here.

\begin{figure}[tbp]
\centering
\includegraphics[width=.4\linewidth]{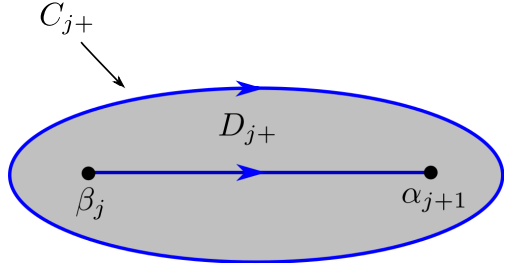}
\caption{\label{Cj-Dj} The contour $C_{j+}$ and the region $D_{j+}$ surrounding $(b_j,a_{j+1})$.}
\end{figure}

Define
\begin{align*}
G(x,t,k) = \begin{mat} e^{-\mc G(x,t,k)} & 0 \\ 0 & e^{\mc G(x,t,k)} \end{mat},
\end{align*}
and the vector-valued function
\begin{align*}
\Sigma(x,t,k) = \varphi(x,t,k) G(x,t,k).
\end{align*}
A direct calculation shows that $\Sigma$ satisfies
\begin{align}\label{sigma-rhp}
 \Sigma^+(x,t,k) &=  \Sigma^-(x,t,k) \begin{choices} \begin{mat} 0 & -e^{-i\Omega_{j+}(x,t)} \\ e^{i\Omega_{j+}(x,t)} & 0 \end{mat}, \when  k \in  (b_j,a_{j+1}),\\
\\
\begin{mat} 0 & -e^{-i\Omega_{j-}(x,t)} \\ e^{i\Omega_{j-}(x,t)} & 0 \end{mat}, \when  k  \in (-a_{j+1},-b_{j}),
\end{choices}
\end{align}
for $j=1,2,\dots,g$ with $\Sigma(x,t,\infty) = [1,1]$. This is a piecewise-constant RHP and we follow ideas from \cite{TrogdonFiniteGenus} to regularize it for numerical purposes.  Define
\begin{align*}
R_{j\pm} (k) &= \half \begin{mat} \beta_{j\pm}(k) + 1/\beta_{\pm j}(k) & i e^{-i\Omega(x,t)} (\beta_{j\pm}(k) - 1/\beta_{j\pm}) \\ -i e^{i\Omega(x,t)} (\beta_{j\pm}(k) - 1/\beta_{j\pm}(k)) & \beta_{j\pm}(k) + 1/\beta_{j\pm}(k) \end{mat},\\
\beta_{j\pm} &= \left( \frac{k\mp a_{j+1}}{k \mp b_j} \right)^{1/4}.
\end{align*}
It follows that $R_{j+}$ ($R_{j-}$) satisfies the same jump as $\Sigma$ in a neighborhood of $(b_j,a_{j+1})$ ($(-a_{j+1},b_j))$.  Let $C_{j+}$ be a clockwise-oriented piecewise-smooth contour lying solely in the right-half plane surrounding $(b_j,a_{j+1})$ but not intersecting or surrounding $(b_i,a_{i+1})$ for $i\neq j$.  Define $C_{j-}$ in an analogous manner for $(-a_{j+1},b_j)$, again with clockwise orientation.  Define $D_{j\pm}$ to be the component of $\mathbb C\setminus C_{j\pm}$ that contains the interval $C_{j\pm}$ encloses, see Figure~\ref{Cj-Dj}.
Define
\begin{align*}
K(x,t,k) = \begin{choices} \Sigma(x,t,k) R_{j\pm}^{-1}(x,t,k), \when k \in D_{j\pm},\\
\Sigma(x,t,k), \otherwise.
\end{choices}
\end{align*}
Then $K(x,t,k)$ solves the following RHP
\begin{align*}
K^+(x,t,k) &= K^-(x,t,k) R_{j\pm}(x,t,k), ~~ k\in C_{j\pm}, ~~ j=1,2,\ldots,g,\\
K(x,t,\infty) &= [1,1].
\end{align*}
This RHP is solved numerically with the method in \cite{SORHFramework} as is seen in \cite{TrogdonFiniteGenus}.

\section{From the $\lambda$-plane to the $k$-plane}\label{Section:Mapping}

We describe a method to transform \eqref{almost-rhp} to an RHP in the $k$-plane so that we may connect it directly with a finite-genus solution of the KdV equation.  First, notice that $\hat \Phi$ fails to be analytic on a subset of $(0,\infty)$. With $k^2 = \lambda$, we write $\hat \Phi(x,t,\lambda) = \chi(x,t,\lambda^{1/2})$ and define
\begin{align*}
\tilde \Phi(x,t,k) = \begin{choices} \chi(x,t,k), \when \imag k > 0,\\
\chi(x,t,-k), \when \imag k < 0. \end{choices}
\end{align*}
It is clear that $\tilde \Phi(k)$ fails to be analytic only on $\mathbb R$.  We compute its jumps.  For $k > 0$
\begin{align*}
\lim_{\epsilon \goto 0^+} \tilde \Phi(x,t,k\pm i\epsilon) = \lim_{\epsilon \goto 0^+} \chi(x,t,k \pm i \epsilon) = \chi^\pm(x,t,k).
\end{align*}
For $k< 0$,
\begin{align*}
\lim_{\epsilon \goto 0^+} \tilde \Phi(x,t,k\pm i\epsilon) = \lim_{\epsilon \goto 0^+} \chi(x,t,-k \mp i \epsilon) = \chi^\mp(x,t,-k).
\end{align*}
For $\lambda > 0$,  if $\hat \Phi^+(x,t,\lambda) = \hat \Phi^-(x,t,\lambda )J(\lambda^{1/2})$ then $\tilde \Phi^+(x,t,k) =  \tilde \Phi^-(x,t,k)J(k)$ for $k > 0$, and $\Phi^-(x,t,k) = \tilde \Phi^+(x,t,k)J(-k)$ for $k < 0$.  Notice that all jumps in \eqref{almost-rhp} satisfy $J(-k) = J^{-1}(k)$.    For ease of notation, define
\begin{align*}
B_+ &= (a_{g+1},\infty) \cup \bigcup_{j=1}^g (a_j,b_j), ~~ B_- =  (-\infty,-a_{g+1}) \cup \bigcup_{j=1}^g (-b_j,-a_j),\\
G_+ &= \bigcup_{j=1}^{g}(b_j,a_{j+1}), ~~ G_- = \bigcup_{j=1}^{g}(-a_{j+1},-b_{j}).
\end{align*}
We are led to an RHP for $\tilde \Phi(x,t,k)$:
\begin{align}\label{tilde-rhp}
\begin{split}
\tilde  \Phi^+(x,t,k) &= \tilde \Phi^-(x,t,k) \begin{mat} 0 & 1 \\ 1 & 0 \end{mat},  ~k \in B_+ \cup B_-,\\
\tilde \Phi^+(x,t,k) &= \tilde \Phi^-(x,t,k) \begin{mat} e^{-2ik x - 8ik^3 t} & 0 \\ 0 & e^{2ik x +8i k^{3} t} \end{mat}, ~ k \in G_+ \cup G_-,\\
\tilde \Phi(x,t,k) &= \begin{mat} 1 & 1 \end{mat}( I + \bigo(1/k)).
\end{split}
\end{align}
Due to its definition, $\tilde \Phi$ solves \eqref{x-jost} in the upper-half plane and the same equation with $k \mapsto -k$ in the lower-half plane.  This leads us to switch the entries of $\tilde \Phi$ in the lower-half plane. Define
\begin{align*}
\tilde \Psi(x,t,k) = \begin{choices} \tilde \Phi(x,t,k), \when \imag k >0,\\
\\
\tilde \Phi(x,t,k) \begin{mat} 0 & 1 \\ 1 & 0 \end{mat}, \when \imag k < 0. \end{choices}
\end{align*}
Thus, $\tilde \Psi(x,t,k)$ satisfies
\begin{align}\label{tilde-psi-rhp}
\begin{split}
\tilde \Psi^+(x,t,k) &= \tilde \Psi^-(x,t,k) \begin{mat} 0 & e^{-2ik x - 8ik^3 t} \\ e^{2ik x +8i k^{3} t} & 0 \end{mat}, ~ k \in G_+ \cup G_-,\\
\tilde \Psi(x,t,k) &= \begin{mat} 1 & 1 \end{mat}( I + \bigo(1/k)).
\end{split}
\end{align}

This differs from the RHP for $\varphi$ given above.  The fundamental difference is that the determinant of the jumps for $\tilde \Psi$ is $-1$ instead of $+1$ in the case of $\varphi$.  As is discussed in \cite{TrogdonFiniteGenus} one column of $\tilde \Psi$ must have a pole in each connected component of $G_+ \cup G_-$. If the pole is at an endpoint of an interval it is a pole on a Riemann surface corresponding to a square-root singularity in the plane.  Given one point from each connected component of $G_+\cup G_-$, it is known that there exists a solution of \eqref{tilde-psi-rhp} that has a pole at each of these points \cite{TrogdonFiniteGenus}.  For the time being, we ignore the presence of poles although they highlight an important issue below.

It follows that we may consider \eqref{tilde-psi-rhp} as a $2 \times 2$ RHP normalized to the identity at infinity.  Summing the rows allows us to obtain a solution of the vector problem as was done in the proof of Theorem~\ref{Theorem:Dressing}.  Consider the auxiliary RHP
\begin{align*}
\nu^+(k) = \nu^-(k) \begin{mat} 0 & -1 \\ 1 & 0 \end{mat}, ~~ k \in G_+ \cup G_-,~~ \nu(\infty) = I.
\end{align*}
Then for
\begin{align*}
\tilde \Psi(x,t,k) = \begin{mat} \tilde \Psi_{11}(x,t,k) & \tilde \Psi_{12}(x,t,k) \\
\tilde \Psi_{21}(x,t,k) & \tilde \Psi_{22}(x,t,k) \end{mat},
\end{align*}
define
\begin{align*}
\tilde \Psi_\nu(x,t,k) = \begin{mat} \nu_{11}(k) \tilde \Psi_{11}(x,t,k) & \nu_{12}(k)\tilde \Psi_{12}(x,t,k) \\
\nu_{21}(k)\tilde \Psi_{21}(x,t,k) & \nu_{22}(k)\tilde \Psi_{22}(x,t,k) \end{mat}.
\end{align*}
A calculation shows that $\tilde \Psi_\nu$ satisfies the same jumps as $\varphi$, see \eqref{delta-rhp}.

It follows that  $\tilde \Psi_\nu$ has a pole in each interval $[b_j,a_{j+1}]$ and $[-a_{j+1},-b_j]$ unless it is precisely cancelled out by an entry of $\nu$.  Thus if we solve the RHP for $\varphi$ and invert the $\tilde \Psi \mapsto \tilde \Psi_\nu$ transformation, we introduce poles at locations determined only by $a_j$ and $b_j$: the zeros of the entries of $\nu$.  Thus this procedure is guaranteed to produce one solution of \eqref{tilde-psi-rhp} despite the fact that there is a whole family of solutions.  This family is described by the fact that for each $\gamma_j \in (b_j,a_{j+1})$ and $\sigma_j \in \{1,2\}$ there exists a solution of \eqref{tilde-psi-rhp} such that $\tilde \Psi_{\sigma_j}$ has a pole at $\pm \gamma_j$.  This is a $g$-parameter family of solutions and it highlights the non-uniqueness of solutions of \eqref{tilde-psi-rhp}.  See \cite{TrogdonFiniteGenus} for details.

\begin{remark}
It follows that $\nu$ can be found explicitly, we follow \cite[p.~281]{FokasPainleve}.  Define
\begin{align*}
\beta(k) = \left( \prod_{j=1}^g \frac{(k-a_{j+1})(k+b_{j})}{(k+a_{j+1})(k-b_j)} \right)^{1/4},
\end{align*}
then
\begin{align*}
\nu(k) = \half \begin{mat} \beta(k) + \beta^{-1}(k) & -i (\beta(k) -\beta^{-1}(k)) \\
i(\beta(k)-\beta^{-1}(k)) & \beta(k) + \beta^{-1}(k) \end{mat}.
\end{align*}
It can be shown that $\beta(k)-\beta^{-1}(k)$ has $2g$ zeros, located at $\pm u_j$ for $u_j \in (b_j,a_{j+1})$ \cite{zhou-notes}.  This justifies the construction above.
\end{remark}

\section{Nonlinear superposition}\label{Section:Superposition}

Below we combine solutions of the IVP with finite-genus solutions using the following definition.
\begin{definition}
Consider two RHPs
\begin{align*}
\Phi_1^+(x,t,k) = \Phi_1^-(x,t,k) e^{-\theta(x,t,k)\hat \sigma_3}V_1(k),~~ k\in \Gamma_1, ~~ \Phi_1(x,t,\infty) = [1,1],\\
\Phi_2^+(x,t,k) = \Phi_2^-(x,t,k) e^{-\theta(x,t,k)\hat \sigma_3}V_2(k),~~ k\in \Gamma_2, ~~ \Phi_2(x,t,\infty) = [1,1],
\end{align*}
such that $V_1$ and $V_2$ satisfy the hypothesis of Theorem~\ref{Theorem:Dressing}.  In addition, assume $V_1$ and $V_2$ commute.  Thus $q_j(x,t) = 2i \lim_{k\goto \infty} k \partial_x \Phi_j(x,t,k)$, $j =1,2$ is a solution of the KdV equation.  We call $q_3(x,t) = 2i \lim_{k\goto \infty} k \partial_x \Phi_3(x,t,k)$ a nonlinear superposition of $q_1(x,t)$ and $q_2(x,t)$ where $\Phi_3(x,t,k)$ solves
\begin{align}\label{Phi3}
\Phi_3^+(x,t,k) = \Phi_3^-(x,t,k) e^{-\theta(x,t,k)\hat \sigma_3}(V_1(k)V_2(k)), ~~k\in \Gamma_1\cup\Gamma_2~~ \Phi_3(x,t,\infty) = [1,1],
\end{align}
and $V_1$ and $V_2$ are extended to be the identity matrix outside their initial domain of definition.
\end{definition}

\begin{remark}
The condition that $V_1$ and $V_2$ commute is necessary so that $V_1V_2$ satisfies the hypotheses of Theorem~\ref{Theorem:Dressing}.
\end{remark}

\begin{example}  
Assume
\begin{align*}
V_1(k) &=\begin{mat} 0 & - 1 \\ 1 & 0 \end{mat}, ~~  k \in [-3,-1] \cup [1,3],\\
V_2(k) &= \begin{mat} 0 & - 1 \\ 1 & 0 \end{mat}, ~~ k \in [-7,-6] \cup [-5,-2] \cup [2,5] \cup [6,7].
\end{align*}
It is trivial that $V_1$ and $V_2$ commute and the corresponding solutions may be superimposed. Here $V_1$ corresponds to a genus-one solution and $V_2$ to a genus-two solution.  Superimposing them produces a new  solution.  The resulting RHP has a jump that is $-I$ on $[-3,-2]$ and $[2,3$. In this way superposition need not happen only when the supports of $V_1-I$ and $V_2-I$ are disjoint.  The symmetries required by the dressing method and the commuting requirement greatly restricts the jumps that can be superimposed.  We only treat the cases where the supports are disjoint.
\end{example}

We make the choice
\begin{align*}
V_1(k) = \begin{mat} 1 - \overline{\rho(\bar k)}\rho(k)  &  - \overline{\rho(\bar k)}\\
\rho(k)& 1 \end{mat},
\end{align*}
where $\rho$ is as in \eqref{FokasLineRHP}.  If $c$ and $\kappa$ are not empty we add additional contours to the RHP. Let
\begin{align*}
V_2(k) = \begin{choices} \begin{mat} 0 & -1 \\ 1 & 0 \end{mat}, \when k \in G_+ \cup G_-,\\
\\
I, \otherwise.
\end{choices}
\end{align*}
We consider the numerical solution of \eqref{Phi3} which represents the nonlinear superposition of the solution of the IVP and a finite-genus solution.

\begin{assumption}
To simplify the computation of solutions, we assume $\rho$ is supported in an interval $[-\ell,\ell]$ and $[-\ell,\ell] \cap (G_+ \cup G_-) = \varnothing$.
\end{assumption}

Thus, we solve the following RHP:
\begin{align*}
\Phi_4^+(k) = \Phi_4^-(k) \begin{choices} e^{-\theta(x,t,k) \hat \sigma_3} V_1(k), \when k \in [-\ell,\ell],\\
e^{-\theta(x,t,k)\hat \sigma_3} V_2(k), \when k \in G_+ \cup G_-. \end{choices}
\end{align*}

\begin{remark}
If $\rho$ has compact support then it certainly cannot be analytic.  In practice, we start with a reflection coefficient $\rho_a$ that is analytic in a strip that contains the real axis.  We construct $\rho$ from $\rho_a$ by multiplying by functions with compact support so that $\rho \approx \rho_a$. This determines $\ell$.  It can be shown using ideas from \cite{TrogdonSONNSD} that the solution $\Phi_a$ of 
\begin{align*}
\Phi_a^+(k) &= \Phi_a^-(k) \begin{choices} e^{-\theta(x,t,k) \hat \sigma_3} \begin{mat} 1- |\rho_a(k)|^2 & -\overline{\rho_a(\bar k )} \\ \rho_a(k) & 1 \end{mat}, \when k \in \mathbb C\setminus(G_+ \cup G_-),\\
\\
e^{-\theta(x,t,k) \hat \sigma_3}\begin{mat} -\overline{\rho_a(\bar k)} - \rho_a(k) & -1 \\ 1 & 0 \end{mat}, \when k \in G_+ \cup G_-, \end{choices}\\
\Phi_a(\infty) &= [1,1],
\end{align*}
 is close to $\Phi_4$ in the sense that if $\|(1+|\cdot|)(\rho(\cdot)-\rho_a(\cdot))\|_{L^1\cap L^\infty(\mathbb R)} < \epsilon$ then $|2i\lim_{|k|\goto\infty}k \partial_x ((\Phi_4)_1-(\Phi_a)_1)| < C \epsilon$, \emph{i.e.}, $2i\lim_{|k|\goto\infty} k\partial_x (\Phi_a)_1$ is a good approximation of the solution of the KdV equation.  Importantly, all the matrix factorizations and contour deformations from \cite{TrogdonSOKdV} can be applied to the RHP for $\Phi_a$ since 
\begin{align*}
\begin{mat} -\overline{\rho_a(\bar k)} - \rho_a(k) & -1 \\ 1 & 0 \end{mat} = \begin{mat} 1 & - \overline{\rho_a(\bar k)} \\ 0 & 1 \end{mat} \begin{mat} 0 & -1 \\ 1 & 0 \end{mat} \begin{mat} 1 & 0 \\ \rho_a(k) & 1 \end{mat}.
\end{align*}

\end{remark}

\begin{figure}[tbp]
\centering
\includegraphics[width=\linewidth]{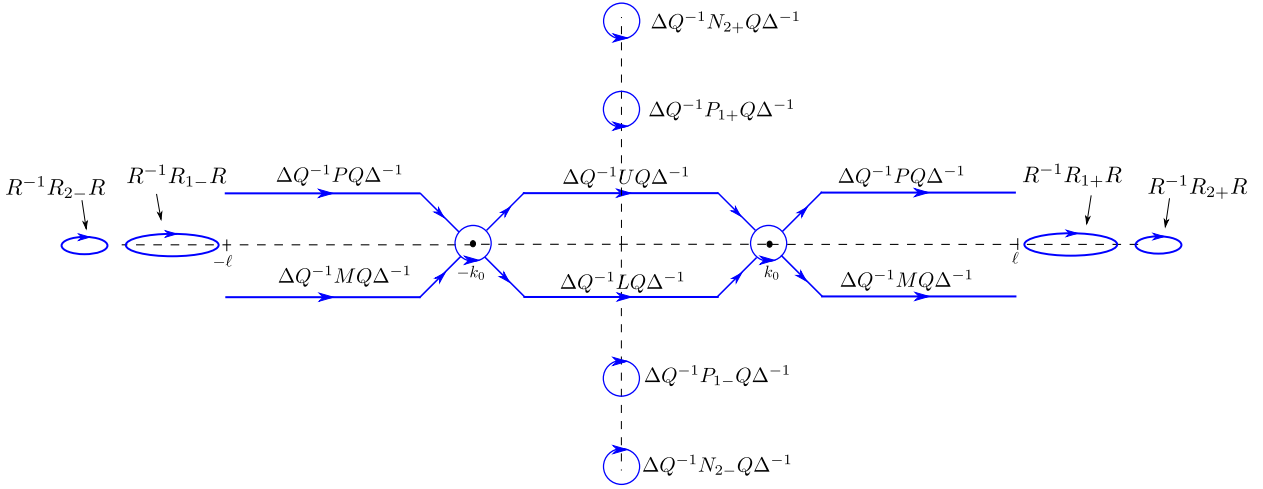}
\caption{\label{Figure:FullRHP} The full RHP that is solved to compute superposition solutions of the KdV equation.  The elliptical contours represent $C_{1\pm}$ and $C_{2\pm}$.  Note that these curves are located outside $[-\ell,\ell]$.}
\end{figure}

The nonlinear steepest descent method as described above transforms $[-\ell,\ell]$ to a contour $\Gamma(x,t)$  with jump $\tilde V_1$ that passes along appropriate paths of steepest descent.  This process affects the jumps on $G_+ \cup G_-$ but only by the multiplication of (to machine precision) analytic, diagonal matrix-valued function $R(x,t,k)$.  The exact form of $R(x,t,k)$ can be inferred from the deformations above.  In the dispersive region $R(x,t,k) = Q^{-1}(k)\Delta(x,t,k)$ and $R(x,t,k) = \Delta(x,t,k)$ for all other regions.  This transforms $V_2(k)$ to  $\tilde V_2(x,t,k) = R^{-1}(x,t,k)V_2(k) R(x,t,k)$.  We display the full RHP for the superposition solutions in Figure~\ref{Figure:FullRHP}.

\begin{remark}
We have highlighted a limitation of our approach.  The contours $C_{j\pm}$ need to be in a location where the reflection coefficient is small.  Furthermore, if $C_{j\pm}$ is near the origin then the corresponding finite-genus solution of the KdV equation has larger period.  Thus, the decay rate of the reflection coefficient affects the periodicity/quasi-periodicity of the finite-genus solution that can be superimposed using this method.
\end{remark}

\section{Numerical Results}\label{Section:Results}

In this section we construct solutions of the KdV equation using the method described above.  We choose a constant $\ell > 0$ and a reflection coefficient $\rho(k)$ for $k \in [-\ell,\ell]$, poles and norming constants ($\kappa = \{\kappa_j\}_{j=1}^n$ and $c=\{c_j\}_{j=1}^n$), and gaps $0 < \ell < b_1 < a_2 < \cdots < a_{g+1}$. 

We note that $q_\pm(x,t)$ in \eqref{sl-fg-time} can be computed. Assume there are $n$ solitons in the solution and for $k_0^2 = -x/(12t) > \ell$  let  $t$ and $x$ be sufficiently large so that $\mathcal K(x,t) = \{1,2,\ldots,n\}$.   Then $R(x,t,k)$ is constant in $x$ and $t$.  Thus the RHP created through the dressing method with $R^{-1}(x,t,k)V_2(k)R(x,t,k)$ defined  on $G_+ \cup G_-$ produces a solution of the KdV equation.  We change the definition of the $g$-function:
\begin{itemize}
\item $\mc G^+(x,t,k) - \mc G^-(x,t,k) = -(2ikx+8ik^3t) - 2 \log R_{11}(x,t,k) + i\Omega_{j+}(x,t)$ for $k  \in (b_j,a_{j+1})$,
\item $\mc G^+(x,t,k) - \mc G^-(x,t,k) = -(2ikx+8ik^3t) - 2 \log R_{11}(x,t,k) + i\Omega_{j-}(x,t)$ for $k  \in (-a_{j+1},-b_j)$.
\end{itemize}
When considering the analog of \eqref{moment-conditions} it is easy to see that the addition of the $\log R_{11}$ term contributes a constant to the right-hand side of the linear system for $\{\Omega_{j\pm}\}_{j=1}^g$.  This induces a phase shift and the effect is shown in plots below.  Note that this modification is not needed for numerical purposes but it highlights the effect of conjugation by $R$.

\subsection{A perturbed genus-two solution with no solitons}

\begin{figure}[tp]
\centering
\includegraphics[width=.8\linewidth]{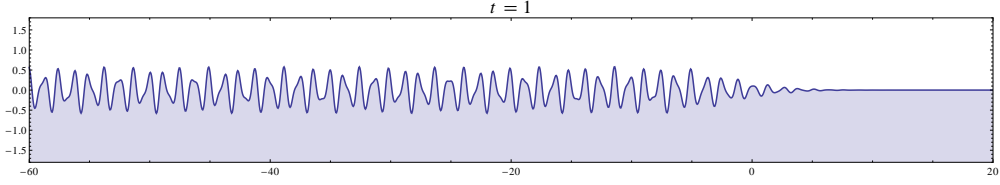}
\caption{\label{Diff-nosoliton} A demonstration of the nonlocal nature of nonlinear superposition: the difference $\tilde q(x,1) = q_1(x,1) + q_2(x,1) - q_3(x,1)$.}
\end{figure}

We choose $\rho$ to be the reflection coefficient obtained from the initial condition $q_0(x) =  -1.2 e^{-(x/4)^2}$ and $\ell =  2.4$.  The sets $c$ and $\kappa$ are both empty.  Finally, we equate $b_1 = 2.5$, $a_2 = 2.54$, $b_2 = 4$ and $a_3 = 4.013$.   Recall that $q_1(x,t)$ is the solution of the KdV equation with initial condition $q_0(x)$, $q_2(x,t)$ is a genus-two solution and $q_3(x,t)$ is the nonlinear superposition.  We present the results in Figures~\ref{D-nosoliton}, \ref{GT-nosoliton} and \ref{DGT-nosoliton} below.  We consider $\tilde q(x,t) = q_1(x,t) + q_2(x,t) -q_3(x,t)$ as a measure of nonlinearity.  See Figure~\ref{Diff-nosoliton} for a plot of $\tilde q(x,t)$ at various times.  We see that the nonlinear interaction is not local: as $x \goto - \infty$ the genus-two solution experiences a phase shift.  Thus the solution obtained from this method is clearly a superposition function for all $t$ in the sense that it satisfies \eqref{sl-fg-time}.

\begin{figure}[tbp]
\centering
\subfigure[]{\includegraphics[width=.8\linewidth]{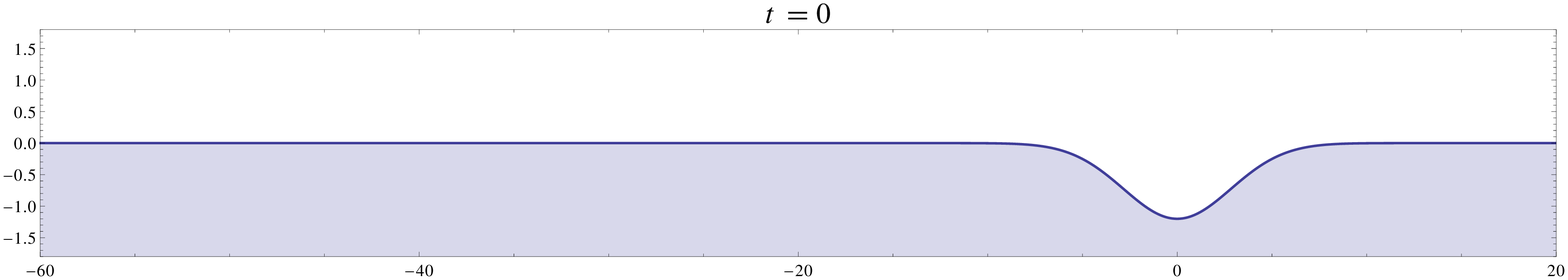}}
\subfigure[]{\includegraphics[width=.8\linewidth]{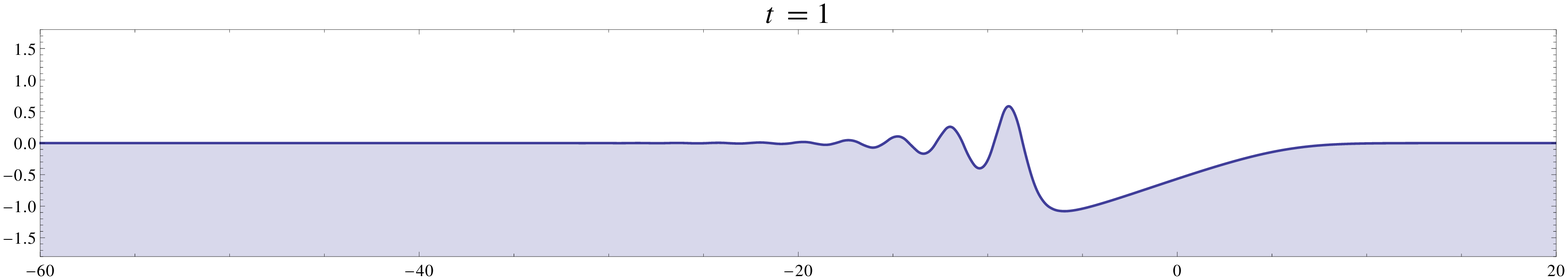}}
\subfigure[]{\includegraphics[width=.8\linewidth]{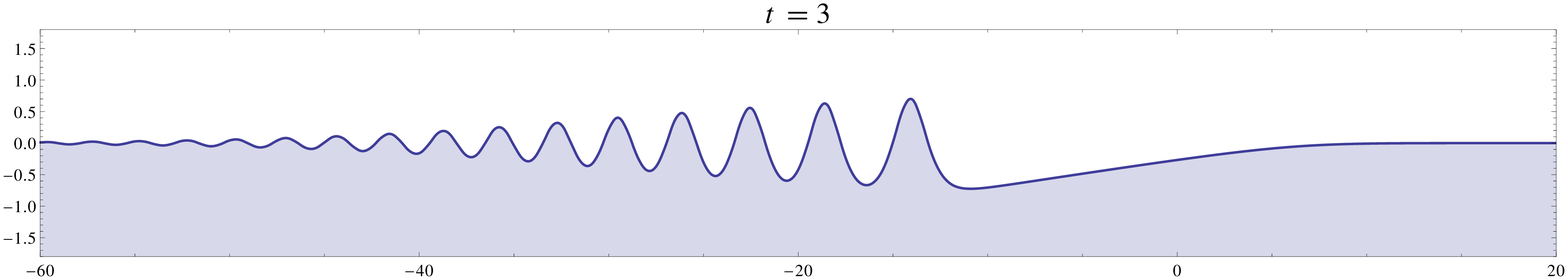}}
\caption{\label{D-nosoliton} (a) The initial condition for $q_1(x,t)$.  (b) A plot of $q_1(x,1)$.  (c) A plot of $q_1(x,3)$.}
\end{figure}

\begin{figure}[tbp]
\centering
\subfigure[]{\includegraphics[width=.8\linewidth]{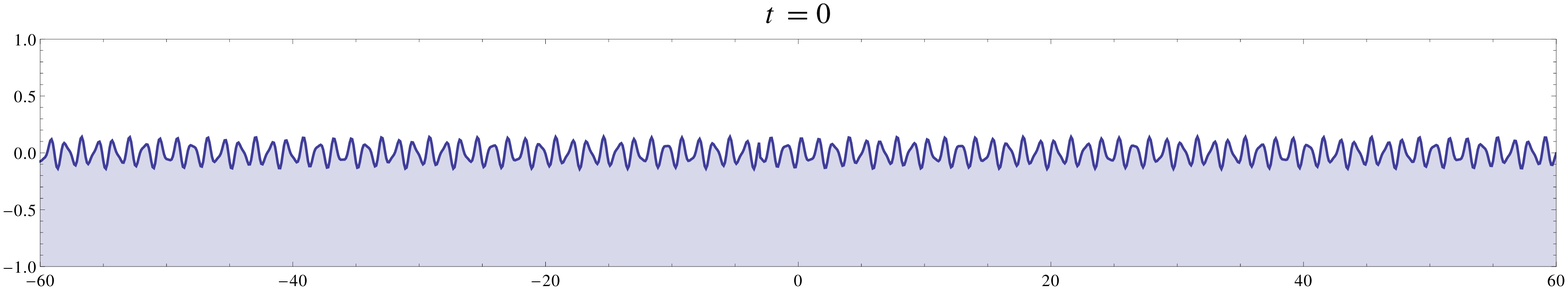}}
\subfigure[]{\includegraphics[width=.5\linewidth]{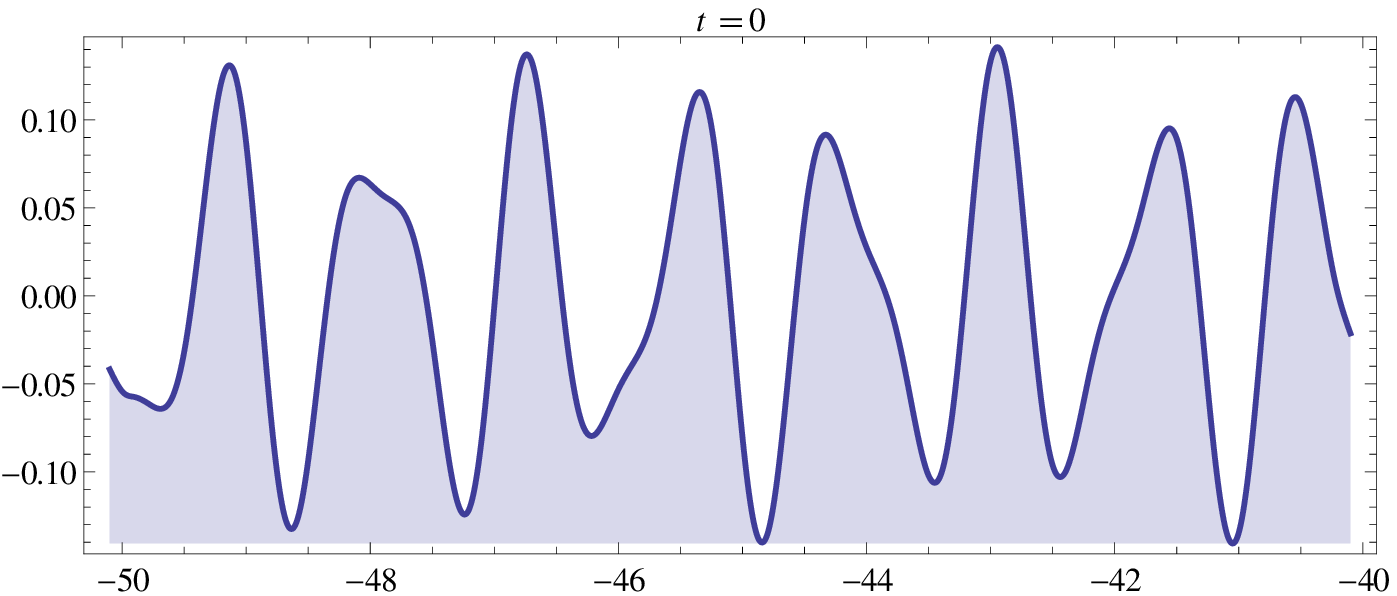}}
\caption{\label{GT-nosoliton} (a) The initial condition for $q_2(x,t)$.  (b) A zoomed plot of $q_2(x,0)$}
\end{figure}

\begin{figure}[tbp]
\centering
\subfigure[]{\includegraphics[width=.8\linewidth]{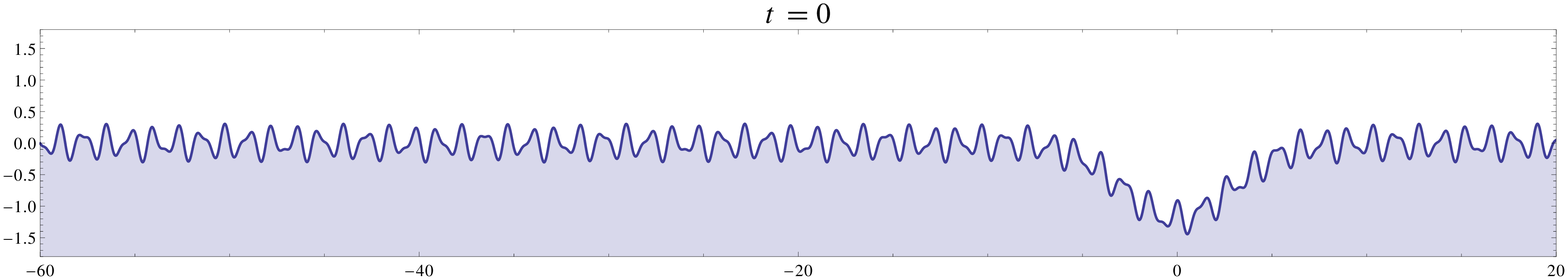}}
\subfigure[]{\includegraphics[width=.8\linewidth]{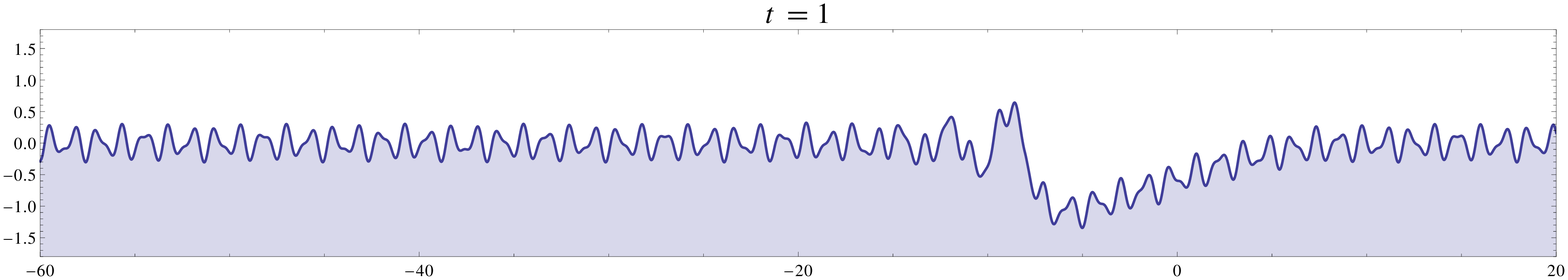}}
\subfigure[]{\includegraphics[width=.8\linewidth]{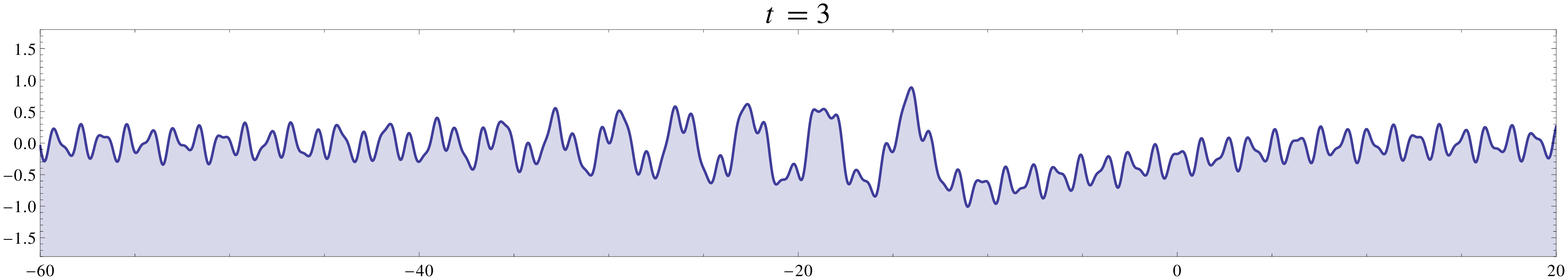}}
\caption{\label{DGT-nosoliton} (a) The initial condition for $q_3(x,t)$.  (b) A plot of $q_3(x,1)$.  (c) A plot of $q_3(x,3)$.}
\end{figure}

\subsection{A perturbed genus-two solution with two solitons}

\begin{figure}[tp]
\centering
\subfigure[]{\includegraphics[width=.8\linewidth]{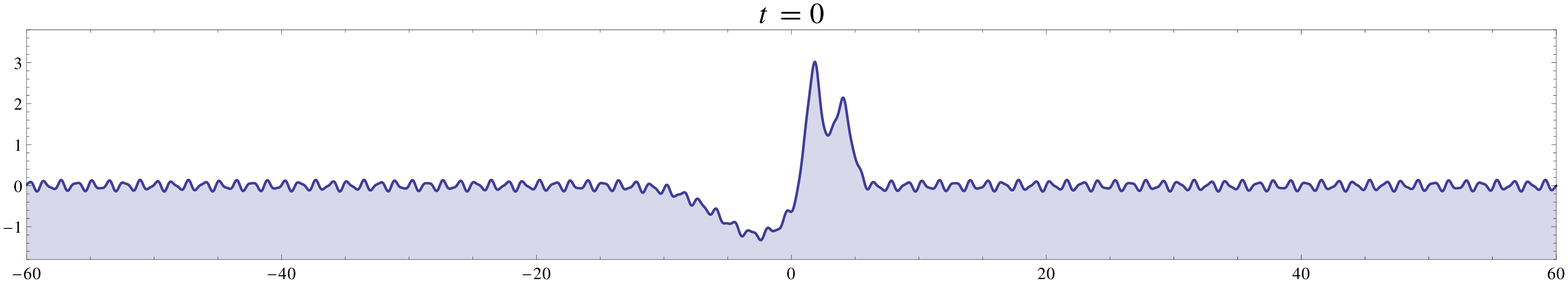}}
\subfigure[]{\includegraphics[width=.8\linewidth]{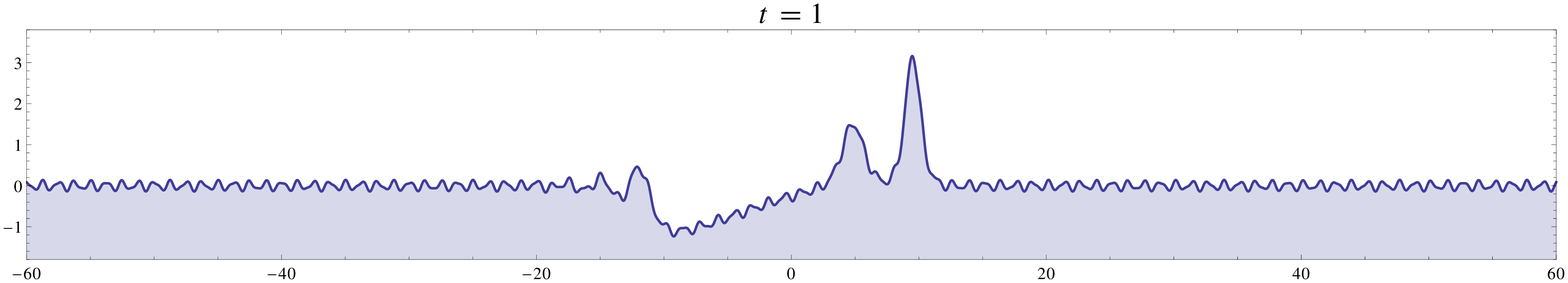}}
\subfigure[]{\includegraphics[width=.8\linewidth]{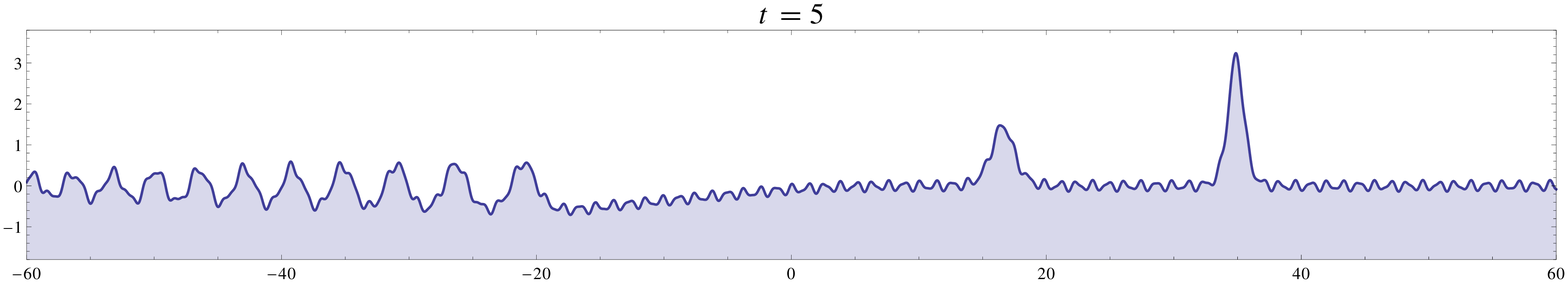}}
\caption{\label{DSGT} The numerical dressing method applied to compute a solution of the KdV equation that contains two solitons, a genus-two solution and dispersion.  (a) The initial condition. (b) A plot of the solution at $t = 1$. (b)  A plot of the solution at $t = 5$.}
\end{figure}

We consider the addition of solitons and dispersion to a genus-two solution.  Again, we let $\rho$ be the reflection coefficient obtained from the initial condition $q_0(x) = -1.2 e^{-(x/4)^2}$.  Also, we choose
\begin{align*}
\kappa = \{1.2589i,0.8571i\}, ~~ c = \{ 7604.0i,1206.3i\}.
\end{align*}
These are chosen by computing the eigenvalues of a positive initial condition.  Finally, to fix the genus-two solution we define $b_1 = 2.5$, $a_2 = 2.52$, $b_2 = 4.1$ and $a_3 = 4.105$.  See Figure~\ref{DSGT} for plots of this solution.

We examine the solution in four regions to demonstrate the phase shifts induced by $R(x,t,k)$ as discussed in the previous sections.  As before, when $R(x,t,k)$ is constant with to its arguments on each component of $G_+ \cup G_-$ we expect the RHP created through the dressing method with $R^{-1}(x,t,k)V_2(k)R(x,t,k)$ defined  on $G_+ \cup G_-$ to produce a genus-two solution of the KdV equation.

\begin{figure}[tp]
\centering
\includegraphics[width=.8\linewidth]{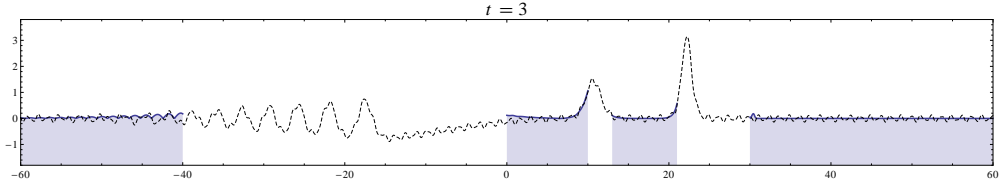}
\caption{\label{Soliton-Diff} A demonstration of the different regions in a two-gap, two-soliton solution. We numerically solve the RHP created through the dressing method with $R^{-1}(x,t,k)V_2(k)R(x,t,k)$ defined  on $G_+ \cup G_-$.  The solution of the KdV equation obtained through this procedure is subtracted from the solution computed from the full RHP (dashed: solution from the full RHP, solid: the absolute difference of the two solutions). In this way we see that the solution limits to a different genus-two solution in each region.}
\end{figure}

These results lead us to the following general conjecture.  When there are no solitons in the solution there are only two regions that are asymptotically close to a finite-genus background: $x \ll 0$ (beyond the dispersive tail) and $x \gg 0$.  With $n$ solitons we have $n+2$ regions:
\begin{itemize}
\item $x \gg 0$ --- in front of all solitons,
\item the $n-1$ regions between solitons,
\item the region between the trailing soliton and the dispersive tail, and
\item $x \ll 0$ --- beyond the dispersive tail.
\end{itemize}
This is consistent with the results of \cite{teschl-asymptotics}.  In Figure~\ref{Soliton-Diff} we demonstrate that using the definition of $R(x,t,k)$ we can compute these solutions.

\section*{Acknowledgments}

We acknowledge the National Science Foundation for its generous support through grant NSF-DMS-1008001 (BD,TT) and NSF-DMS-1303018 (TT).  Any opinions, findings, and conclusions or recommendations expressed in this material are those of the authors and do not necessarily reflect the views of the funding sources.

\bibliographystyle{plain}
\bibliography{Dressing}

\def\cprime{$'$} \def\cprime{$'$} \def\cprime{$'$}
  \def\cydot{\leavevmode\raise.4ex\hbox{.}}
  \def\cydot{\leavevmode\raise.4ex\hbox{.}} \def\cprime{$'$}
\begin{thebibliography}{10}

\bibitem{ablowitz-segur-book}
M.~Ablowitz and H.~Segur.
\newblock {\em Solitons and the Inverse Scattering Transform}.
\newblock SIAM, Philadelpha, PA, 1981.

\bibitem{AblowitzClarksonSolitons}
M.~J. Ablowitz and P.~A. Clarkson.
\newblock {\em Solitons, Nonlinear Evolution Equations and Inverse Scattering}.
\newblock Cambridge University Press, 1991.

\bibitem{AblowitzSegurSolution}
M.~J. Ablowitz and H.~Segur.
\newblock Asymptotic solutions of the {Korteweg--de Vries} equation.
\newblock {\em Stud. in Appl. Math.}, 57:13--44, 1977.

\bibitem{belokolos1}
E.~D. Belokolos, A.~I. Bobenko, V.~Z. Enol'skii, A.~R. Its, and V.~B. Matveev.
\newblock {\em Algebro-geometric approach to nonlinear integrable problems}.
\newblock Springer Series in Nonlinear Dynamics. Springer-Verlag, Berlin, 1994.

\bibitem{deconinckpatterson}
B.~Deconinck and M.~S. Patterson.
\newblock Computing with plane algebraic curves and {R}iemann surfaces: the
  algorithms of the {M}aple package ``algcurves''.
\newblock In {\em Computational approach to {R}iemann surfaces}, volume 2013 of
  {\em Lecture Notes in Math.}, pages 67--123. Springer, Heidelberg, 2011.

\bibitem{DeiftOrthogonalPolynomials}
P.~Deift.
\newblock {\em {Orthogonal Polynomials and Random Matrices: a Riemann-Hilbert
  Approach}}.
\newblock AMS, 2000.

\bibitem{Deift-gfun}
P.~Deift, S.~Venakides, and X.~Zhou.
\newblock An extension of the steepest descent method for {R}iemann-{H}ilbert
  problems: the small dispersion limit of the {K}orteweg-de {V}ries ({K}d{V})
  equation.
\newblock {\em Proc. Natl. Acad. Sci. USA}, 95:450--454, 1998.

\bibitem{DeiftZhouAMS}
P.~Deift and X.~Zhou.
\newblock A steepest descent method for oscillatory {Riemann--Hilbert}
  problems.
\newblock {\em Bulletin AMS}, 26:119--124, 1992.

\bibitem{deift-zhou-ven}
P.~Deift, X.~Zhou, and S.~Venakides.
\newblock The collisionless shock region for the long-time behavior of
  solutions of the {K}d{V} equation.
\newblock {\em Comm. Pure and Appl. Math.}, 47:199--206, 1994.

\bibitem{Dressing}
E.~V. Doktorov and S.~B. Leble.
\newblock {\em A dressing method in mathematical physics}, volume~28 of {\em
  Mathematical Physics Studies}.
\newblock Springer, Dordrecht, 2007.

\bibitem{Dubrovin}
B.~A. Dubrovin.
\newblock Inverse problem for periodic finite zoned potentials in the theory of
  scattering.
\newblock {\em Func. Anal. and Its Appl.}, 9:61--62, 1975.

\bibitem{dubrovin75}
B.~A. Dubrovin.
\newblock The inverse scattering problem for periodic finite-zone potentials.
\newblock {\em Funkcional. Anal. i Prilo\v zen.}, 9(1):65--66, 1975.

\bibitem{DubrovinTheta}
B.~A. Dubrovin.
\newblock Theta functions and non-linear equations.
\newblock {\em Russian Math. Surveys}, 36:11--92, 1981.

\bibitem{teschl-finite-gap}
I.~Egorova, K.~Grunert, and G.~Teschl.
\newblock On the {C}auchy problem for the {K}orteweg-de {V}ries equation with
  steplike finite-gap initial data. {I}. {S}chwartz-type perturbations.
\newblock {\em Nonlinearity}, 22(6):1431--1457, 2009.

\bibitem{FokasUnified}
A.~S. Fokas.
\newblock {\em A Unified Approach to Boundary Value Problems}.
\newblock SIAM, Philadelphia, PA, 2008.

\bibitem{FokasPainleve}
A.~S. Fokas, A.~R. Its, A.~A. Kapaev, and V.~Y. Novokshenov.
\newblock {\em {Painlev\'e} Transcendents: the Riemann--Hilbert Approach}.
\newblock AMS, 2006.

\bibitem{frauendienerklein}
J.~Frauendiener and C.~Klein.
\newblock Algebraic curves and {R}iemann surfaces in {M}atlab.
\newblock In {\em Computational approach to {R}iemann surfaces}, volume 2013 of
  {\em Lecture Notes in Math.}, pages 125--162. Springer, Heidelberg, 2011.

\bibitem{teschl}
K.~Grunert and G.~Teschl.
\newblock Long-time asymptotics for the {Korteweg--de Vries} equation via
  nonlinear steepest descent.
\newblock {\em Math. Phys., Anal. and Geom.}, 12:287--324, 2008.

\bibitem{itsmatveev1}
A.~R. Its and V.~B. Matveev.
\newblock Hill operators with a finite number of lacunae.
\newblock {\em Funkcional. Anal. i Prilo\v zen.}, 9(1):69--70, 1975.

\bibitem{itsmatveev2}
A.~R. Its and V.~B. Matveev.
\newblock Schr\"odinger operators with the finite-band spectrum and the
  {$N$}-soliton solutions of the {K}orteweg-de {V}ries equation.
\newblock {\em Teoret. Mat. Fiz.}, 23(1):51--68, 1975.

\bibitem{matveev}
V.~B. Matveev.
\newblock 30 years of finite-gap integration theory.
\newblock {\em Philos. Trans. R. Soc. Lond. Ser. A Math. Phys. Eng. Sci.},
  366(1867):837--875, 2008.

\bibitem{mckeanvanmoerbeke}
H.~P. McKean and P.~{v}an Moerbeke.
\newblock The spectrum of {H}ill's equation.
\newblock {\em Invent. Math.}, 30(3):217--274, 1975.

\bibitem{teschl-asymptotics}
A.~Mikikits-Leitner and G.~Teschl.
\newblock Long-time asymptotics of perturbed finite-gap {K}orteweg-de {V}ries
  solutions.
\newblock {\em J. Anal. Math.}, 116:163--218, 2012.

\bibitem{TheoryOfSolitons}
S.~Novikov, S.~V. Manakov, L.~P. Pitaevskii, and V.~E. Zakharov.
\newblock {\em Theory of Solitons}.
\newblock Constants Bureau, New York, 1984.

\bibitem{novikov}
S.~P. Novikov.
\newblock A periodic problem for the {K}orteweg-de {V}ries equation. {I}.
\newblock {\em Funkcional. Anal. i Prilo\v zen.}, 8(3):54--66, 1974.

\bibitem{SOPainleveII}
S.~Olver.
\newblock Numerical solution of {Riemann--Hilbert} problems: {Painlev\'e II}.
\newblock {\em Found. Comput. Math.}, 2010.

\bibitem{SORHFramework}
S.~Olver.
\newblock A general framework for solving {R}iemann-{H}ilbert problems
  numerically.
\newblock {\em Numer. Math.}, 122(2):305--340, 2012.

\bibitem{TrogdonSONNSD}
S.~Olver and T.~Trogdon.
\newblock Nonlinear steepest descent and the numerical solution of
  {Riemann--Hilbert} problems.
\newblock {\em to appear in Comm. Pure Appl. Math.}, 2012.

\bibitem{TrogdonFiniteGenus}
T.~Trogdon and B.~Deconinck.
\newblock A {Riemann--Hilbert} problem for the finite-genus solutions of the
  {KdV} equation and its numerical solution.
\newblock {\em to appear in Physica D.}, 2012.

\bibitem{TrogdonSOKdV}
T.~Trogdon, S.~Olver, and B.~Deconinck.
\newblock Numerical inverse scattering for the {Korteweg-de Vries} and modified
  {Korteweg-de Vries} equations.
\newblock {\em Physica D}, 241:1003--1025, 2012.

\bibitem{ZakharovDressing}
V.~E. Zakharov.
\newblock On the dressing method.
\newblock In {\em Inverse methods in action ({M}ontpellier, 1989)}, Inverse
  Probl. Theoret. Imaging, pages 602--623. Springer, Berlin, 1990.

\bibitem{zhou-RHP}
X.~Zhou.
\newblock The {Riemann--Hilbert} problem and inverse scattering.
\newblock {\em SIAM J. Math. Anal.}, 20:966--986, 1989.

\bibitem{zhou-notes}
X.~Zhou.
\newblock {Riemann--Hilbert} problems and integrable systems.
\newblock {\em Lectures at MSRI}, 1999.

\end{thebibliography}

\end{document}